\documentclass[12pt,letterpaper]{article}
\usepackage{epsfig,rotating,setspace,latexsym,amsmath,epsf,amssymb,amsfonts,bm,theorem,cite,algorithm,algorithmic,graphicx,epsf,authblk,epstopdf,color,accents,amssymb,bbm}

\newtheorem{assumption}{Assumption}

\newtheorem{corollary}{Corollary}
\newtheorem{definition}{Definition}
\newtheorem{theorem}{Theorem}
\newtheorem{lemma}{Lemma}
\newenvironment{Proof}[1]{\medskip\par\noindent{\bf Proof:\,}\,#1}{{\mbox{\,$\blacksquare$}\par}}

\allowdisplaybreaks

\begin{document}

\title{Age-Minimal Transmission for Energy Harvesting Sensors with Finite Batteries: Online Policies\thanks{Ahmed Arafa and H. Vincent Poor are with the Electrical Engineering Department, Princeton University, NJ 08544. Jing Yang is with the School of Electrical Engineering and Computer Science, The Pennsylvania State University, PA 16802. Sennur Ulukus is with the Department of Electrical and Computer Engineering, University of Maryland College Park, MD 20742. Emails: \emph{aarafa@princeton.edu}, \emph{yangjing@psu.edu}, \emph{ulukus@umd.edu}, \emph{poor@princeton.edu}. This work is presented in part in the 2018 Information Theory and Applications Workshop \cite{arafa-age-sgl}, and in the 2018 International Conference on Communications \cite{arafa-age-rbr}.}}

\author{Ahmed Arafa, Jing Yang, Sennur Ulukus, and~H. Vincent Poor}

\maketitle

%\vspace{-.5in}

%============
\begin{abstract}
An energy-harvesting sensor node that is sending status updates to a destination is considered. The sensor is equipped with a battery of {\it finite} size to save its incoming energy, and consumes one unit of energy per status update transmission, which is delivered to the destination instantly over an error-free channel. The setting is {\it online} in which the harvested energy is revealed to the sensor causally over time after it arrives, and the goal is to design status update transmission times (policy) such that the long term average {\it age of information} (AoI) is minimized. The AoI is defined as the time elapsed since the latest update has reached at the destination. Two energy arrival models are considered: a {\it random battery recharge} (RBR) model, and an {\it incremental battery recharge} (IBR) model. In both models, energy arrives according to a Poisson process with unit rate, with values that completely fill up the battery in the RBR model, and with values that fill up the battery incrementally in a unit-by-unit fashion in the IBR model. The key approach to characterizing the optimal status update policy for both models is showing the optimality of {\it renewal policies}, in which the inter-update times follow a renewal process in a certain manner that depends on the energy arrival model and the battery size. It is then shown that the optimal renewal policy has an energy-dependent {\it threshold} structure, in which the sensor sends a status update only if the AoI grows above a certain threshold that depends on the energy available in its battery. For both the random and the incremental battery recharge models, the optimal energy-dependent thresholds are characterized {\it explicitly}, i.e., in closed-form, in terms of the optimal long term average AoI. It is also shown that the optimal thresholds are monotonically decreasing in the energy available in the battery, and that the smallest threshold, which comes in effect when the battery is full, is equal to the optimal long term average AoI.
\end{abstract}

%============
\section{Introduction}

Real-time sensing applications in which time-sensitive measurement status updates of some physical phenomenon are sent to a destination (receiver) calls for careful transmission scheduling policies under proper metrics that assess the updates' timeliness and freshness. The {\it age of information} (AoI) metric has recently acquired attention as a suitable candidate for such a purpose. The AoI is defined as the time spent since the latest measurement update has reached the destination, and hence it basically captures delay from the destination's perspective. When sensors (transmitters) rely on energy harvested from nature to transmit their status updates, they cannot transmit all the time, so that they do not run out of energy and risk having overly stale status updates at the destination. Therefore, the fundamental question as to how to optimally manage the harvested energy to send timely status updates needs to be addressed. In this work, we provide an answer to this question by deriving {\it optimal} status update policies for energy harvesting sensors with {\it finite} batteries in an {\it online} setting where the harvested energy is only revealed causally over time.

The online energy harvesting communication literature, in which energy arrival information is only revealed causally over time, is mainly studied via Markov decision processes modeling and dynamic programming techniques, see, e.g.,  \cite{omurFade, ruiZhangEH, sharma-online-eh, mao-online-eh, mitran-eh-online, dong-eh-online, elif-eh-online}, and also via specific analyses of the involved stochastic processes, as in \cite{koksal-online-eh, jing-online-collab-sense, jing-online-sensing}. A different approach is introduced in \cite{ozgur_online_su}, and then extended in \cite{ozgur_online_mac, baknina_online_mac, baknina_online_bc, baknina-online-proc, varan-online-ffp-cognitive, baknina-online-data, ozgur-online-gen, arafa-baknina-ffp, arafa-mobile-tgcn, ozgur-online-block-iid} for various system models, in which an online fixed fraction policy, where transmitters use a fixed fraction of their available energy for transmission in each time slot, is shown to perform within a constant gap from the optimal online policy. Such fixed fraction online policies are simpler than usual online policies introduced in the literature, with provable near-optimal performance. In the online setting of this work, we also investigate a relatively simple online policy, and show its {\it exact optimality}.

The AoI metric has been studied in the literature under various settings; mainly through modeling the update system as a queuing system and analyzing the long term average AoI, and through using optimization tools to characterize optimal status update policies, see, e.g., \cite{yates_age_1, yates_age_mac, ephremides_age_random, ephremides_age_management, ephremides_age_non_linear, shroff_age_multi_hop, modiano-age-bc, soljanin-age-multicast, najm-age-multistream, sun-age-mdp, yates-age-erase-code, yates-age-cache, sun-weiner}, and also the recent survey in \cite{kosta-age-monograph}. In this work, we employ tools from optimization theory to devise age-minimal online status update policies in systems where sensors are energy-constrained, and rely on energy harvested from nature to transmit status updates. Some related works to this problem are summarized next.

AoI analysis and optimization in energy harvesting communications have been recently considered in \cite{yates_age_eh, elif_age_eh, liu-age-eh-sensing, arafa-age-2hop, arafa-age-var-serv, elif-age-Emax, jing-age-online, jing-age-error-infinite-no-fb, jing-age-error-infinite-w-fb, baknina-age-coding, shahab-age-online-rndm, baknina-updt-info} under various service time (time for the update to take effect), battery capacity, and channel assumptions. With the exception of \cite{arafa-age-var-serv}, an underlying assumption in these works is that energy expenditure is normalized, i.e., sending one status update consumes one energy unit. References \cite{yates_age_eh, elif_age_eh} consider a sensor with infinite battery, with \cite{yates_age_eh} focusing on online policies under stochastic service times, and \cite{elif_age_eh} focusing on both offline and online policies with zero service times, i.e., with updates reaching the destination instantly. Reference \cite{liu-age-eh-sensing} studies the effect of sensing costs on AoI with an infinite battery sensor transmitting through a noisy channel. Using a harvest-then-use protocol, \cite{liu-age-eh-sensing} presents a steady state analysis of AoI under both deterministic and random energy arrivals.
The offline policy in \cite{elif_age_eh} is extended to non-zero, but fixed, service times in \cite{arafa-age-2hop} for both single and multi-hop settings, and in \cite{arafa-age-var-serv} for energy-controlled variable service times. The online policy in \cite{elif_age_eh} is analyzed through a dynamic programming approach in a discretized time setting, and is shown to have a threshold structure, i.e., an update is sent only if the age grows above a certain threshold and energy is available for transmission. Motivated by such results for the infinite battery case, \cite{elif-age-Emax} then studies the performance of online threshold policies for the finite battery case under zero service times, yet with no proof of optimality. Reference \cite{jing-age-online} proves the optimality of online threshold policies under zero service times for the special case of a unit-sized battery, via tools from renewal theory. It also shows the optimality of best effort online policies, where updates are sent over uniformly-spaced time intervals if energy is available, for the infinite battery case. Such best effort online policy is also shown to be optimal, for the infinite battery case, when updates are subject to erasures in \cite{jing-age-error-infinite-no-fb, jing-age-error-infinite-w-fb}; with no erasure error feedback in \cite{jing-age-error-infinite-no-fb} and with perfect erasure error feedback in \cite{jing-age-error-infinite-w-fb}. Under the same system model of \cite{jing-age-error-infinite-no-fb}, reference \cite{baknina-age-coding} analyzes the performances of the best effort online policy and the save-and-transmit online policy, where the sensor saves some energy in its battery before attempting transmission, under coding to combat channel erasures. A slightly different system model is considered in \cite{shahab-age-online-rndm}, in which status updates are externally arriving, i.e., their measurement times are not controlled by the sensor. With a finite battery, and stochastic service times, reference \cite{shahab-age-online-rndm} employs tools from stochastic hybrid systems to analyze the long term average AoI. An interesting approach is followed in \cite{baknina-updt-info} where the idea of sending extra information, on top of the measurement status updates, is introduced and analyzed for unit batteries and zero service times.

In this paper, we show the {\it optimality of online threshold policies} under a {\it finite battery} setting, with zero service times. We consider two energy arrival (recharging) models, namely, a {\it random battery recharge} (RBR) model, and an {\it incremental battery recharge} (IBR) model. In both models, energy arrives according to a Poisson process with unit rate, yet with the following difference: in the RBR model, energy arrivals {\it completely} fill up the battery, while in the IBR model, energy arrivals fill up the battery incrementally in a {\it unit-by-unit} fashion. We invoke tools from renewal theory to show that the optimal status update policy, the one minimizing the long term average AoI, is such that specific update times, depending on the recharging model, follow a {\it renewal} process with independent inter-update delays. Then, we follow a Lagrangian approach to show that the optimal renewal-type policy, for both recharging models, has an {\it energy-dependent threshold} structure, in which an update is sent only if the AoI grows above a certain threshold that depends on the energy available in the battery, the specifics of which vary according to the recharging model. Our approach enables characterizing the optimal thresholds {\it explicitly,} i.e., in closed-form, in terms of the optimal long term average AoI, which is in turn found by a bisection search over an interval that is strictly smaller than the unit interval. We also show that, for both recharging models, the optimal thresholds are monotonically decreasing in the available energy, i.e., the higher the available energy the smaller the corresponding threshold, and that the smallest threshold, corresponding to a full battery, is equal to the optimal long term average AoI.

We acknowledge an independent and concurrent work \cite{elif-age-online-threshold} that considers the same setting of the IBR model considered in this work, and also shows the optimality of online threshold policies. In there, tools from the theory of optimal stopping, from the stochastic control literature, are invoked to show such result, along with some structural properties. The optimal thresholds are found numerically. Different from \cite{elif-age-online-threshold}, however, and as mentioned above, the approach followed in this work for the IBR model, namely, proving the renewal structure of the optimal policy followed by the Lagrangian approach, allows characterizing the optimal energy-dependent thresholds in closed-form in terms of the optimal AoI.

%============
\section{System Model and Problem Formulation}

We consider a sensor node that collects measurements from a physical phenomenon and sends updates to a destination over time. The sensor relies on energy harvested from nature to acquire and send its measurement updates, and is equipped with a battery of finite size $B$ to save its incoming energy. The sensor consumes one unit of energy to measure and send out an update to the destination. We assume that updates are sent over an error-free link with negligible transmission times as in \cite{elif_age_eh, elif-age-Emax, jing-age-online}. Energy arrives (is harvested) at times $\{t_1,t_2,\dots\}$ according to a Poisson process of rate $1$. Our setting is online in which energy arrival times are revealed causally over time; only the arrival rate is known a priori. We consider two models for the amount of harvested energy at each arrival time. The first model, denoted {\it random battery recharge} (RBR), is when energy arrives in $B$ units. This models, e.g., situations where the battery size is relatively small with respect to the amounts of harvested energy, and hence energy arrivals fully recharge the battery. We note that this RBR model has been previously considered in the online scheduling literature in \cite{ozgur_online_su, ozgur_online_mac, baknina_online_mac, baknina_online_bc, baknina-online-proc, varan-online-ffp-cognitive, baknina-online-data, ozgur-online-gen, arafa-baknina-ffp, arafa-mobile-tgcn, ozgur-online-block-iid} and in the information-theoretic approach considered in \cite{ozgur_rbr}. The second model, denoted {\it incremental battery recharge} (IBR), is when energy arrives in units, i.e., when the battery is recharged incrementally in a unit-by-unit fashion. We mathematically illustrate the difference between the two models below.

Let $s_i$ denote the time at which the sensor acquires (and transmits) the $i$th measurement update, and let $\mathcal{E}(t)$ denote the amount of energy in the battery at time $t$. We then have the following energy causality constraint \cite{jingP2P}:
\begin{align} \label{eq_en_caus}
\mathcal{E}\left(s_i^-\right)\geq1,\quad\forall i.
\end{align}
We assume that we begin with an empty battery at time $0$. For the RBR model, the battery evolves as follows over time:
\begin{align} \label{eq_battery_rbr}
\mathcal{E}\left(s_i^-\right)=\min\left\{\mathcal{E}\left(s_{i-1}^-\right)-1+ B\cdot\mathcal{A}\left(x_i\right),B\right\},
\end{align}
where $x_i\triangleq s_i-s_{i-1}$, and $\mathcal{A}(x_i)$ denotes the number of energy arrivals in $[s_{i-1},s_i)$. Note that $\mathcal{A}(x_i)$ is a Poisson random variable with parameter $x_i$. We denote by $\mathcal{F}_B$ the set of feasible transmission times $\{s_i\}$ described by (\ref{eq_en_caus}) and (\ref{eq_battery_rbr}) in addition to an empty battery at time 0, i.e., $\mathcal{E}(0)=0$. Similarly, for the IBR model, the battery evolves as follows over time:
\begin{align} \label{eq_battery_inc}
\mathcal{E}\left(s_i^-\right)=\min\left\{\mathcal{E}\left(s_{i-1}^-\right)-1+\mathcal{A}\left(x_i\right),B\right\}.
\end{align}
We denote by $\mathcal{F}$, the set of feasible transmission times $\{s_i\}$ described by (\ref{eq_en_caus}) and (\ref{eq_battery_inc}) in addition to an empty battery at time 0, i.e., $\mathcal{E}(0)=0$.

\begin{figure}[t]
\center
\includegraphics[scale=1]{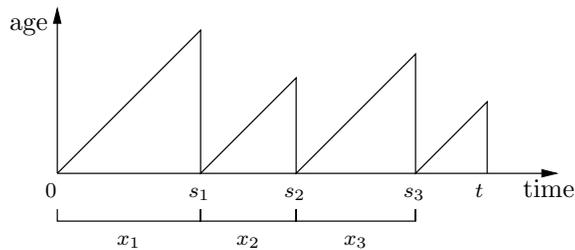}
\caption{Example of the age evolution versus time with $n(t)=3$.}
\label{fig_age_xmpl}
\end{figure}

For either recharging model, the goal is to choose an online feasible transmission policy $\{s_i\}$ (or equivalently $\{x_i\}$) such that the long term average of the AoI experienced at the destination is minimized. The AoI is defined as the time elapsed since the latest update has reached the destination, which is formally defined as follows at time $t$:
\begin{align}
a(t)\triangleq t-u(t),
\end{align}
where $u(t)$ is the time stamp of the latest update received before time $t$. Let $n(t)$ denote the total number of updates sent by time $t$. We are interested in minimizing the area under the age curve representing the total cumulative AoI, see Fig.~\ref{fig_age_xmpl} for a possible sample path with $n(t)=3$. At time $t$, this area is given by
\begin{align} \label{eq_aoi}
r(t)\triangleq\frac{1}{2}\sum_{i=1}^{n(t)}x_i^2+\frac{1}{2}\left(t-s_{n(t)}\right)^2,
\end{align}
and therefore the goal is to characterize the following quantity for the RBR model:
\begin{align} \label{opt_main_rbr}
\rho_B\triangleq\min_{{\bm x}\in\mathcal{F}_B}\limsup_{T\rightarrow\infty}\frac{1}{T}\mathbb{E}\left[r(T)\right]
\end{align}
representing the long term average AoI, where $\mathbb{E}(\cdot)$ is the expectation operator. Similarly, for the IBR model, the goal is to characterize
\begin{align} \label{opt_main_inc}
\rho\triangleq\min_{{\bm x}\in\mathcal{F}}\limsup_{T\rightarrow\infty}\frac{1}{T}\mathbb{E}\left[r(T)\right].
\end{align}
We discuss problems (\ref{opt_main_rbr}) and (\ref{opt_main_inc}) in Sections~\ref{sec_rbr} and \ref{sec_inc}, respectively. In the next section, we discuss the special case of $B=1$ in which the two models are equivalent.

%============
\section{Unit Battery Case: A Review}

In this section, we review the case $B=1$ studied in \cite{jing-age-online}.  Observe that for $B=1$, $\mathcal{F}_B=\mathcal{F}$ and problems (\ref{opt_main_rbr}) and (\ref{opt_main_inc}) are identical. In studying this problem, reference \cite{jing-age-online} first shows that renewal policies, i.e., policies with update times $\{s_i\}$ forming a renewal process, outperform any other {\it uniformly bounded policy}, which are defined as follows (see \cite[Definition 3]{jing-age-online}).
\begin{definition}[Uniformly Bounded Policy] \label{def_ubp}
An online policy whose inter-update times, as a function of the energy arrival times, have a bounded second moment.
\end{definition}
Then, reference \cite{jing-age-online} shows that the optimal renewal policy is a threshold policy, where an update is sent only if the AoI grows above a certain threshold. We review this latter result in this section.

Let $\tau_i$ denote the time until the next energy arrival since the $(i-1)$th update time, $s_{i-1}$. Since the arrival process is Poisson with rate 1, $\tau_i$'s are independent and identically distributed (i.i.d.) exponential random variables with parameter 1. Under renewal policies, the $i$th inter-update time $x_i$ should not depend on the events before $s_{i-1}$; it can only be a function of $\tau_i$. Moreover, under any feasible policy, $x_i(\tau_i)$ cannot be smaller than $\tau_i$, since the battery is empty at $s_{i-1}$. Next, note that whenever an update occurs, both the battery and the age drop to 0, and hence the system resets. This constitutes a renewal event, and therefore using the strong law of large numbers of renewal processes \cite{ross_stochastic}, problem (\ref{opt_main_rbr}) (or equivalently problem (\ref{opt_main_inc})) reduces to
\begin{align} \label{opt_b1}
\rho_1=\min_{x(\tau)\geq\tau}\quad&\frac{\mathbb{E}\left[x(\tau)^2\right]}{2\mathbb{E}[x(\tau)]},
\end{align}
where expectation is over the exponential random variable $\tau$.

In order to make problem (\ref{opt_b1}) more tractable to solve, we introduce the following parameterized problem:
\begin{align} \label{opt_b1_lmda}
p_1(\lambda)\triangleq\min_{x(\tau)\geq\tau}\quad&\frac{1}{2}\mathbb{E}\left[x(\tau)^2\right]-\lambda\mathbb{E}[x(\tau)].
\end{align}
This approach was discussed in \cite{dinkelbach-fractional-prog}. We now have the following lemma, which is a restatement of the results in \cite{dinkelbach-fractional-prog}, and provide a proof for completeness.

\begin{lemma} \label{thm_lmda}
$p_1(\lambda)$ is decreasing in $\lambda$, and the optimal solution of problem (\ref{opt_b1}) is given by $\lambda^*$ that solves $p_1(\lambda^*)=0$.
\end{lemma}

\begin{Proof}
Let $\lambda_1>0$, and let the solution of problem (\ref{opt_b1_lmda}) be given by $x^{(1)}$ for $\lambda=\lambda_1$. Now for some $\lambda_2>\lambda_1$, one can write
\begin{align}
p_1(\lambda_1)&=\frac{1}{2}\mathbb{E}\left[\left(x^{(1)}\right)^2\right]-\lambda_1\mathbb{E}\left[x^{(1)}\right] \nonumber \\
&>\frac{1}{2}\mathbb{E}\left[\left(x^{(1)}\right)^2\right]-\lambda_2\mathbb{E}\left[x^{(1)}\right] \nonumber \nonumber \\
&\geq p_1(\lambda_2),
\end{align}
where the last inequality follows since $x^{(1)}$ is also feasible in problem (\ref{opt_b1_lmda}) for $\lambda=\lambda_2$.

Next, note that both problems (\ref{opt_b1_lmda}) and (\ref{opt_b1}) have the same feasible set. In addition, if $p_1(\lambda)=0$, then the objective function of (\ref{opt_b1}) satisfies $\frac{1}{2}\mathbb{E}\left[\left(x^{(1)}\right)^2\right]/\mathbb{E}\left[x^{(1)}\right]=\lambda$. Hence, the objective function of (\ref{opt_b1}) is minimized by finding the minimum $\lambda\geq0$ such that $p_1(\lambda)=0$. Finally, by the first part of lemma, there can only be one such $\lambda$, which we denote $\lambda^*$.
\end{Proof}

By Lemma~\ref{thm_lmda}, one can simply use a bisection method to find $\lambda^*$ that solves $p_1(\lambda^*)=0$. This $\lambda^*$ certainly exists since $p_1(0)>0$ and $\lim_{\lambda\rightarrow\infty}p_1(\lambda)=-\infty$. We focus on problem (\ref{opt_b1_lmda}) in the rest of this section, for which we introduce the following Lagrangian \cite{boyd}:
\begin{align}
\mathcal{L}=&\frac{1}{2}\int_0^\infty x^2(\tau)e^{-\tau}d\tau-\lambda\int_0^\infty x(\tau)e^{-\tau}d\tau -\int_0^\infty\mu(\tau)\left(x(\tau)-\tau\right)d\tau,
\end{align}
where $\mu(\tau)$ is a non-negative Lagrange multiplier. Taking derivative with respect to $x(t)$ and equating to 0 we get
\begin{align} \label{eq_x_b1_kkt}
x(t)=\lambda+\frac{\mu(t)}{e^{-t}}.
\end{align}
Now if $t<\lambda$, then $x(t)$ has to be larger than $t$, for if it were equal, the right hand side of the above equation would be larger than the left hand side. By complementary slackness \cite{boyd}, we conclude that in this case $\mu(t)=0$, and hence $x(t)=\lambda$. On the other hand if $t\geq\lambda$, then $x(t)$ has to be equal to $t$, for if it were larger, then by complementary slackness $\mu(t)=0$ and the right hand side of the above equation would be smaller than the left hand side. In conclusion, we have
\begin{align}
x(t)=\begin{cases}\lambda,\quad &t\leq\lambda\\
t,\quad &t>\lambda\end{cases}.
\end{align}

This means that the optimal inter-update time is threshold-based; if an energy arrival occurs before $\lambda$ amount of time since the last update time, i.e., if $\tau<\lambda$, then the sensor should not use this energy amount right away to send an update. Instead, it should wait for $\lambda-\tau$ extra amount of time before updating. Else, if an energy arrival occurs after $\lambda$ amount of time since the last update time, i.e., if $\tau\geq\lambda$, then the sensor should use that amount of energy to send an update right away. We coin this kind of policy {\it $\lambda$-threshold policy}. Substituting this $x(t)$ into problem (\ref{opt_b1_lmda}) we get
\begin{align} \label{eq_p1_lmda}
p_1(\lambda)=e^{-\lambda}-\frac{1}{2}\lambda^2,
\end{align}
which admits a unique solution of $\lambda^*\approx0.9012$ when equated to 0. In the next two sections, we extend the above approach to characterize optimal policies for larger (general) battery sizes under both RBR and IBR models.

%============
\section{Random Battery Recharge (RBR) Model} \label{sec_rbr}

\subsection{Renewal-Type Policies}

In this section, we focus on problem (\ref{opt_main_rbr}) in the general case of $B>1$ energy units. Let $l_i$ denote the $i$th time that the battery level falls down to $B-1$ energy units. We use the term {\it epoch} to denote the time duration between two consecutive such events, and define $x_{B,i}\triangleq l_i-l_{i-1}$ as the length of the $i$th epoch. The main reason behind choosing such specific event to determine the epoch's start/end times is that the epoch would then contain at most $B$ updates, and that any other choice leads to having possibly infinite number of updates in a single epoch, which is clearly more complex to analyze. Let $\tau_i$ denote the time until the next energy arrival after $l_{i-1}$.  One scenario for the update process in the $i$th epoch would be that starting at time $l_{i-1}$, the sensor sends an update only after the battery recharges, i.e., at some time after $l_{i-1}+\tau_i$, causing the battery state to fall down from $B$ to $B-1$ again. Another scenario would be that the sensor sends $j\leq B-1$ updates before the battery recharges, i.e., at some times before $l_{i-1}+\tau_i$, and then submits one more update after the recharge occurs, making in total $j+1$ updates in the $i$th epoch.

\begin{figure}[t]
\center
\includegraphics[scale=1]{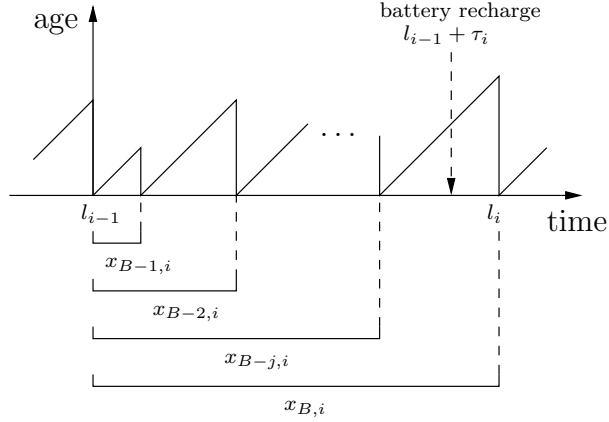}
\caption{Age evolution over time in the $i$th epoch, with $j+1\leq B$ updates.}
\label{fig_age_ex}
\end{figure}

Let us now define $x_{j,i}$, $1\leq j\leq B-1$, to be the time it takes the sensor to send $B-j$ updates in the $i$th epoch before a battery recharge occurs. That is, starting at time $l_{i-1}$, and assuming that the $i$th epoch contains $B$ updates, the sensor sends the first update at $l_{i-1}+x_{B-1,i}$, followed by the second update at $l_{i-1}+x_{B-2,i}$, and so on, until it submits the $B-1$st update at $l_{i-1}+x_{1,i}$, using up all the energy in its battery. The sensor then waits until it gets a recharge at $l_{i-1}+\tau_i$ before sending its final $B$th update in the epoch. See Fig.~\ref{fig_age_ex} for an example run of the AoI curve during the $i$th epoch given that the sensor sends $j+1\leq B$ updates.

In general, under any feasible status updating online policy, $\{x_{j,i}\}_{j=1}^{B-1}$ and $x_{B,i}$ may depend on all the history of status updating and energy arrival information up to $l_{i-1}$, which we denote by $\mathcal{H}_{i-1}$. In addition to that, the value of $x_{B,i}$ can also depend on $\tau_i$. However, by the energy causality constraint (\ref{eq_en_caus}), the values of $\{x_{j,i}\}_{j=1}^{B-1}$ cannot depend on $\tau_i$. This is due to the fact that if the sensor updates $j+1$ times in the same epoch, then the first $j$ updates should occur before the battery recharges. Focusing on uniformly bounded policies, we now have the following theorem. The proof is in Appendix~\ref{apndx_rbr_rnwl}.

\begin{theorem} \label{thm_rbr_rnwl}
The optimal status update policy for problem (\ref{opt_main_rbr}) in the case $B>1$ is a renewal policy, i.e., the sequence $\{l_i\}$ forms a renewal process. Moreover, the optimal $\{x_{j,i}\}_{j=1}^{B-1}$ are constants, and the optimal $x_{B,i}$ only depends on $\tau_i$.
\end{theorem}

\subsection{Threshold Policies}

Theorem~\ref{thm_rbr_rnwl} indicates that the sensor should let its battery level fall down to $B-1$ at times that constitute a renewal policy. Next, we characterize the optimal renewal policy by which the sensor sends its updates. Using the strong law of large numbers of renewal processes (renewal reward theorem) \cite[Theorem~3.6.1]{ross_stochastic}, problem (\ref{opt_main_rbr}) reduces to an optimization over a single epoch as follows:
\begin{align} \label{opt_bk}
\rho_B=\min_{{\bm x}}\quad&\frac{\mathbb{E}\left[R\left({\bm x}\right)\right]}{\mathbb{E}\left[x_B(\tau)\right]} \nonumber \\
\mbox{s.t.}\quad&x_{B-1}\geq0 \nonumber\\
&x_{j-1}\geq x_j,\quad 2\leq j\leq B-1 \nonumber \\
&x_B(\tau)\geq\tau,\quad\forall\tau,
\end{align}
where ${\bm x}\triangleq\{x_1,\dots,x_B\}$, with $x_B(t)$ denoting the length of an epoch in which the battery recharge occurs after $t$ time units of its beginning, and $R\left({\bm x}\right)$ denotes the area under the age curve during an epoch. Note that the expectation is over the exponential random variable $\tau$. Using the renewal-reward theorem enables one, by the i.i.d. property of epochs, to consider optimizing the status update policy over a single epoch, and then repeat it over all other epochs, without losing optimality. This is the main essence of problem (\ref{opt_bk}). Similar to the $B=1$ case, we define $p_B(\lambda)$ as follows:
\begin{align} \label{opt_bk_lmda}
p_B^{\text{rbr}}(\lambda) \triangleq \min_{{\bm x}} \quad &\mathbb{E}\left[R\left({\bm x}\right)\right]-\lambda\mathbb{E}\left[x_B(\tau)\right] \nonumber \\
\mbox{s.t.} \quad &\text{constraints of (\ref{opt_bk})}.
\end{align}
As in Lemma~\ref{thm_lmda}, one can show that $p_B^{\text{rbr}}(\lambda)$ is decreasing in $\lambda$, and that the optimal solution of problem (\ref{opt_bk}) is given by $\lambda^*$ satisfying $p_B^{\text{rbr}}(\lambda^*)=0$.

\begin{figure}[t]
\center
\includegraphics[scale=1]{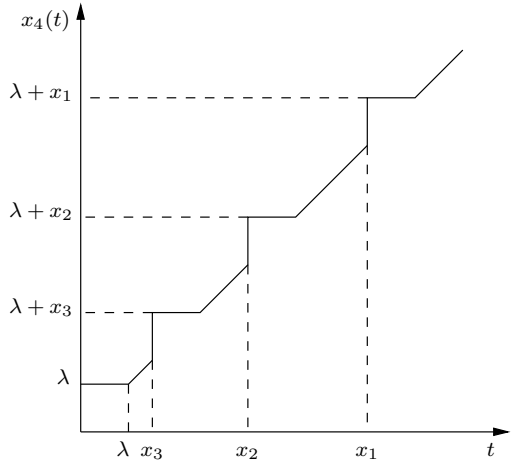}
\caption{Optimal multi threshold structure of $x_4(t)$.}
\label{fig_th_policy}
\end{figure}

Since the optimal solution for the $B>1$ case cannot be larger than that of the $B=1$ case, which is $0.9012$, one can use, e.g., a bisection search over $(0,0.9012]$ to find the optimal $\lambda$ for $B>1$. We now write the following Lagrangian for problem (\ref{opt_bk_lmda}) after expanding the objective function:
\begin{align}
\mathcal{L}=&\frac{1}{2}x_{B-1}^2e^{-x_{B-1}}+\frac{1}{2}\sum_{j=1}^{B-2}\left(x_j-x_{j+1}\right)^2e^{-x_j} +\frac{1}{2}\int_0^{x_{B-1}}x_B(\tau)^2e^{-\tau}d\tau \nonumber\\
&+\frac{1}{2}\sum_{j=2}^{B-1}  \int_{x_j}^{x_{j-1}}  \left(x_B(\tau)-x_j\right)^2e^{-\tau}d\tau  +\frac{1}{2}\int_{x_1}^\infty\left(x_B(\tau)-x_1\right)^2e^{-\tau}d\tau -  \lambda  \int_0^\infty  x_B(\tau)e^{-\tau}d\tau \nonumber\\
& -\mu_{B-1}x_{B-1} -\sum_{j=1}^{B-2}\mu_j\left(x_j-x_{j+1}\right) -\int_0^\infty\mu_B(\tau)\left(x_B(\tau)-\tau\right)d\tau,
\end{align}
where $\left\{\mu_1,\dots,\mu_{B-1},\mu_B(\tau)\right\}$ are non-negative Lagrange multipliers. Taking derivative with respect to $x_B(t)$ and equating to 0 we get
\begin{align}
x_B(t)=\lambda+\sum_{j=1}^{B-1}x_j\mathbbm{1}_{x_j\leq t<x_{j-1}}+\frac{\mu_B(t)}{e^{-t}},
\end{align}
where $\mathbbm{1}_A$ equals 1 if the event $A$ is true, and 0 otherwise. Now let us assume that $\lambda$ is smaller than $\min\left\{x_{B-1},\min_{1\leq j\leq B-2}x_j-x_{j+1}\right\}$, and verify this assumption later on. Proceeding similarly to the analysis of the $B=1$ case, we get
\begin{align} \label{eq_xk_opt}
x_B(t)=\begin{cases}\lambda,\quad&t<\lambda\\t,\quad&\lambda\leq t<x_{B-1}\\\lambda+x_{B-1},\quad&x_{B-1}\leq t<\lambda+x_{B-1}\\t,\quad&\lambda+x_{B-1}\leq t<x_{B-2}\\
\vdots\\
\lambda+x_1,\quad&x_1\leq t<\lambda+x_1\\
t,\quad&t\geq\lambda+x_1\end{cases}.
\end{align}
A depiction of the above policy for $B=4$ is shown in Fig.~\ref{fig_th_policy}.

Thus, the optimal update policy has the following structure. Starting with a battery of $B-1$ energy units and zero age, if the next battery recharge occurs at any time before $\lambda$ time units, then the sensor updates at exactly $t=\lambda$. While if it occurs at any time between $\lambda$ and $x_{B-1}$, then the sensor updates right away. This is the same as the {\it $\lambda$-threshold policy}, the solution of the $B=1$ case, except that it has a cut-off at $t=x_{B-1}$. This cut-off value has the following interpretation: if the battery recharge does not occur until $t=x_{B-1}$, then the sensor updates at $t=x_{B-1}$, causing the battery level and the age to fall down to $B-2$ and $0$, respectively. The sensor then repeats the {\it $\lambda$-threshold policy} described above with a new cut-off value of $x_{B-2}$, i.e., if the recharge does not occur until $t=x_{B-2}$, then the sensor updates again at $t=x_{B-2}$, causing the battery level and the age to fall down to $B-3$ and $0$, respectively. This technique repeats up to $t=x_1$, when the sensor updates for the $(B-1)$th time, emptying its battery. At this time, the sensor waits for the battery recharge and applies the {\it $\lambda$-threshold policy} one last time, with no cut-off value, to submit the last $B$th update in the epoch. Note that if the battery recharge occurs at some time $\tau<x_{B-1}$, then there would be $1$ update in the epoch. On the other hand, if $x_j\leq\tau<x_{j-1}$, for some $2\leq j\leq B-1$, then there would be $B-j+1$ updates. Finally, if $\tau\geq x_1$ then there would be $B$ updates.

In the sequel, we find the optimal values of $\{x_j\}_{j=1}^{B-1}$ (and $\lambda$) by taking derivatives of the Lagrangian with respect to $x_j$'s and equating to 0. Before doing so, we simplify the objective function of problem (\ref{opt_bk_lmda}) by evaluating the expectations involved and using (\ref{eq_xk_opt}). After some simplifications we get
\begin{align} \label{eq_obj_fn}
&\mathbb{E}\left[R\left({\bm x}\right)\right]-\lambda\mathbb{E}\left[x_B(\tau)\right]= \nonumber\\
&\qquad\qquad e^{-\lambda}-\frac{1}{2}\lambda^2+f_1(\lambda)\sum_{j=1}^{B-1}e^{-x_j} \left(x_{B-1}+1\right)e^{-x_{B-1}}-\sum_{j=1}^{B-2}\left(x_j-x_{j+1}+1\right)e^{-x_j},
\end{align}
where $f_1(\lambda)\triangleq\lambda+e^{-\lambda}-\frac{1}{2}\lambda^2$. Using the above in the Lagrangian and taking derivatives we get
\begin{align}
x_1&=x_2+f_1(\lambda)+\mu_1e^{x_1}, \label{eq_kkt_x1} \\
x_j&=x_{j+1}+f_1(\lambda)+\left(\mu_j-\mu_{j-1}-e^{x_{j-1}}\right)e^{-x_j},\quad 2\leq j\leq B-2, \label{eq_kkt_xj} \\
x_{B-1}&=f_1(\lambda)+\left(\mu_{B-1}-\mu_{B-2}-e^{-x_{B-2}}\right)e^{x_{B-1}}. \label{eq_kkt_xk1}
\end{align}
Now let us assume that $x_j>x_{j+1}$, $1\leq j\leq B-2$, and $x_{B-1}>0$. Hence, by complementary slackness we have $\mu_j=0$, $1\leq j\leq B-1$. One can then substitute $x_1-x_2$ in (\ref{eq_kkt_xj}) for $j=2$ to find $x_2-x_3$ and proceed recursively to get
\begin{align}
x_j-x_{j+1}&=f_j(\lambda),\quad 1\leq j\leq B-2, \label{eq_xj_lmda}\\
x_{B-1}&=f_{B-1}(\lambda), \label{eq_xk1_lmda}
\end{align}
where we have defined
\begin{align} \label{eq_rbr_fj}
f_j(\lambda)&\triangleq f_1(\lambda)-e^{-f_{j-1}(\lambda)},\quad2\leq j\leq B-1.
\end{align}
We have the following result on the structure of $\{f_j(\lambda)\}$.

\begin{lemma} \label{thm_fj_dec}
For a fixed $\lambda$, the sequence $\{f_j(\lambda)\}_{j=1}^{B-1}$ is decreasing; and for a fixed $j$, $f_j(\lambda)$ is decreasing in $\lambda$.
\end{lemma}

\begin{Proof}
The proofs of the two statements follow by induction. Clearly, we have $f_2(\lambda)<f_1(\lambda)$. Now assume $f_j(\lambda)<f_{j-1}(\lambda)$ for some $j>2$. Therefore $f_{j+1}(\lambda)=f_1(\lambda)-e^{-f_j(\lambda)}<f_1(\lambda)-e^{-f_{j-1}(\lambda)}=f_j(\lambda)$. This shows the first statement.

Next, direct first derivative analysis shows that $f_1(\lambda)$ is decreasing in $\lambda$. Now assume that $f_j(\lambda)$ is decreasing in $\lambda$ for some $j\geq2$, and observe that $\frac{df_{j+1}(\lambda)}{d\lambda}=\frac{df_1(\lambda)}{d\lambda}+e^{-f_j(\lambda)}\frac{df_j(\lambda)}{d\lambda}$, which is negative by the induction hypothesis. This shows the second statement, and completes the proof of the lemma.
\end{Proof}

%\begin{lemma} \label{thm_fj_dec}
%For a fixed $\lambda$, the sequence $\{f_j(\lambda)\}_{j=1}^{k-1}$ is decreasing; for a fixed $j$, $f_j(\lambda)$ is decreasing in $\lambda$; and $f_j(0)>0$, $1\leq j\leq k-1$.
%\end{lemma}
%\begin{Proof}
%The proofs of the three statements follow by induction. Clearly, we have $f_2(\lambda)<f_1(\lambda)$. Now assume $f_j(\lambda)<f_{j-1}(\lambda)$. Therefore $f_{j+1}(\lambda)=f_1(\lambda)-e^{-f_j(\lambda)}<f_1(\lambda)-e^{-f_{j-1}(\lambda)}=f_j(\lambda)$. This shows the first statement.
%
%Next, direct first derivative analysis shows that $f_1(\lambda)$ is decreasing in $\lambda$. Now assume that $f_j(\lambda)$ is decreasing in $\lambda$, and observe that $\frac{df_{j+1}(\lambda)}{d\lambda}=\frac{df_1(\lambda)}{d\lambda}+e^{-f_j(\lambda)}\frac{df_j(\lambda)}{d\lambda}$, which is negative by the induction hypothesis. This shows the second statement.
%
%Finally, note that $f_1(0)=1$. Now assume $f_j(0)>0$. Therefore, $f_{j+1}(0)=f_1(0)-e^{-f_j(0)}=1-\frac{1}{e^{f_j(0)}}>0$. This shows the third statement, and completes the proof of the lemma.
%\end{Proof}

Note that $f_j(\lambda)$ represents the inter-update delay between updates $B-j-1$ and $B-j$. With this in mind, Lemma~\ref{thm_fj_dec} has an intuitive explanation: it shows that when the amount of energy in the battery is relatively low, the sensor becomes less eager to send the next update, so that it does not run out of energy, and oppositely, when the amount of energy in the battery is relatively high, the sensor becomes more eager to send the next update so that it makes use of the available energy before the next recharge overflows the battery. Next, by equations (\ref{eq_xj_lmda}) and (\ref{eq_xk1_lmda}), we proceed recursively from $j=B-1$ to $j=1$ to find the values of $x_j$'s in terms of $\lambda$. This gives
\begin{align} \label{eq_thresholds_rbr}
x_j=\sum_{m=j}^{B-1}f_m(\lambda),\quad 1\leq j\leq B-1.
\end{align}
Finally, we substitute the above in (\ref{eq_obj_fn}) to get 
\begin{align} 
p_B^{\text{rbr}}(\lambda)=&e^{-\lambda}-\frac{1}{2}\lambda^2+\sum_{j=1}^{B-1}\left(f_1(\lambda)-f_j(\lambda)-1\right)e^{-\sum_{m=j}^{B-1}f_j(\lambda)} \\
=&e^{-\lambda}-\frac{1}{2}\lambda^2-e^{-f_{B-1}(\lambda)}, \label{eq_pk_lmda}
\end{align}
and perform a bisection search over $\lambda\in(0,0.9012]$ to find the optimal $\lambda^*$ that solves $p_B^{\text{rbr}}(\lambda^*)=0$. We note that for $B=1$, the summation in (\ref{eq_pk_lmda}) vanishes and we directly get (\ref{eq_p1_lmda}). Finally, observe that $p_B^{\text{rbr}}(\lambda)=0$ implies $f_{B-1}(\lambda)=-\log\left(e^{-\lambda}-\frac{1}{2}\lambda^2\right)$. Since $0<\lambda\leq0.9012$, we have $0\leq e^{-\lambda}-\frac{1}{2}\lambda^2<1$, and hence $f_{B-1}(\lambda)>0$; moreover $f_{B-1}(\lambda)>-\log\left(e^{-\lambda}\right)=\lambda$. By Lemma~\ref{thm_fj_dec}, the above argument shows that: 1) $f_j\left(\lambda^*\right)>0$, $1\leq j\leq B-1$, which further implies by (\ref{eq_kkt_x1})-(\ref{eq_kkt_xk1}) that all Lagrange multipliers are zero, as previously assumed; and 2) $\lambda^*<f_j\left(\lambda^*\right)$, $1\leq j\leq B-1$, which verifies the previous assumption regarding the optimal age being smaller than all inter-update delays.

In summary, given the functions $\{f_j(\lambda)\}_{j=2}^{B-1}$ through the recursive formulas in (\ref{eq_rbr_fj}) with $f_1(\lambda)=\lambda+e^{-\lambda}-\frac{1}{2}\lambda^2$, the optimal solution of problem (\ref{opt_main_rbr}) is given by a bisection search for $\lambda^*$ that satisfies $p_B^{\text{rbr}}(\lambda^*)=0$ in (\ref{eq_pk_lmda}), and the thresholds $\{x_j^*\}_{j=1}^{B-1}$ of the optimal policy in (\ref{eq_xk_opt}) are given by (\ref{eq_thresholds_rbr}).

%============
\section{Incremental Battery Recharge (IBR) Model} \label{sec_inc}

\subsection{Renewal-Type Policies}

In this section, we focus on problem (\ref{opt_main_inc}) in the general case of $B>1$. Similar to what we have shown in the previous section, we first show that the optimal update policy that solves problem (\ref{opt_main_inc}) has a renewal structure. Namely, we show that it is optimal to transmit updates in such a way that the inter-update delays are independent over time; and that the time durations in between the two consecutive events of transmitting an update and having $k\leq B-1$ units of energy left in the battery are i.i.d., i.e., these events occur at times that constitute a renewal process. We first introduce some notation.

Let the pair $\left(\mathcal{E}(t),a(t)\right)$ represent the state of the system at time $t$. Fix $k\in\{0,1,\dots,B-1\}$, and consider the state $(k,0)$, which means that the sensor has just submitted an update and has $k$ units of energy remaining in its battery. Let $l_i$ denote the time at which the system visits $(k,0)$ for the $i$th time. We use the term {\it epoch} to denote the time in between two consecutive visits to $(k,0)$. Observe that there can possibly be an infinite number of updates occurring in an epoch, depending on the energy arrival pattern and the update time decisions. For instance, in the $i$th epoch, which starts at $l_{i-1}$, one energy unit may arrive at some time $l_{i-1}+\tau_{1,i}$, at which the system goes to state $(k+1,\tau_{1,i})$, and then the sensor updates afterwards to get the system state back to $(k,0)$ again. Another possibility (if $k\geq1$) is that the sensor first updates at some time $l_{i-1}+x_{k,i}$, at which the system goes to state $(k-1,0)$, and then two consecutive energy units arrive at times $l_{i-1}+\tau_{1,i}$ and $l_{i-1}+\tau_{1,i}+\tau_{2,i}$, respectively, at which the system goes to state $(k+1,\tau_{1,i}+\tau_{2,i})$, and then the sensor updates afterwards to get the system state back to $(k,0)$ again. Depending on how many energy arrivals occur in the $i$th epoch, how far apart from each other they are, and the status update times, one can determine the length of the $i$th epoch and how many updates it has. Observe that the update policy in the $i$th epoch may depend on the history of events (energy arrivals and transmission updates) that occurred in previous epochs, which we denote by $\mathcal{H}_{i-1}$. Our first main result in this section shows that this is not the case, under uniformly bounded policies as per Definition~\ref{def_ubp}, and that epoch lengths should be i.i.d. Our next theorem formalizes this. The proof is in Appendix~\ref{apndx_inc_rnwl}.

\begin{theorem} \label{thm_inc_rnwl}
The optimal status update policy for problem (\ref{opt_main_inc}) in the case $B>1$ is a renewal policy, i.e., the sequence $\{l_i\}$ denoting the times at which the system visits state $(k,0)$, for some fixed $0\leq k\leq B-1$, forms a renewal process.
\end{theorem}

Based on Theorem~\ref{thm_inc_rnwl}, the next corollary now follows.

\begin{corollary} \label{thm_inc_indp}
In the optimal solution of problem (\ref{opt_main_inc}), the inter-update times are independent.
\end{corollary}

\begin{Proof}
Observe that whenever an update occurs the system enters state $(j,0)$ for some $j\leq B-1$. The system then starts a new epoch {\it with respect to state} $(j,0)$. Since the choice of $k$ energy units in Theorem~\ref{thm_inc_rnwl} is arbitrary, the results of the theorem now tell us that the update policy in that epoch, and therefore its length, is independent of the past history, in particular the past inter-update lengths.
\end{Proof}

Based on Corollary~\ref{thm_inc_indp}, we have the following observation. Let us assume that the optimal policy is such that the state at time $t$ is $(j,\tau)$. This means that the previous status update occurred at time $t-\tau$. By Corollary~\ref{thm_inc_indp}, the policy at time $t$ is independent of the events before time $t-\tau$. However, it may depend on the events occurring in $[t-\tau,t)$. For instance, for $j\geq1$, it may be the case that at time $(t-\tau)^+$ the sensor had $j-1$ energy units in its battery, and then received another energy unit at some time in $[t-\tau,t)$; or, it may have already started with $j$ energy units at time $(t-\tau)^+$ and received no extra energy units in $[t-\tau,t)$. The question now is whether the optimal policy at time $t$ is the same in either of the two scenarios. The following result concludes that it is indeed the same.

\begin{lemma} \label{thm_state_dep_policy}
The optimal status update policy of problem (\ref{opt_main_inc}) is such that at time $t$ the next scheduled update time is only a function of the system state $(\mathcal{E}(t),a(t))$.
\end{lemma}

\begin{Proof}
Let us assume that the optimal policy is such that the state at time $t$ is $(j,\tau)$. Then this means that the previous status update occurred at time $t-\tau$. By Corollary~\ref{thm_inc_indp}, the optimal policy at time $t$ in this case is independent of the events before $t-\tau$. Starting from time $t$, the sensor then solves a {\it shifted} problem defined as follows. We basically use the same terminology and random variables that constitute (\ref{eq_aoi}) to characterize the area under the age curve starting from time $t$ until time $t+T$ (instead of starting from time $0$ to time $T$), and denote it by $r_t(T)$, with $a(t)=\tau$. We also characterize a shifted feasible set $\mathcal{F}_t$, in which the battery evolves exactly as in (\ref{eq_battery_inc}) and starts with $j$ energy units at time $t$. Therefore, given a state of $(j,\tau)$ at time $t$, the sensor solves the following shifted problem:
\begin{align}
\min_{{\bm x}\in\mathcal{F}_t}\limsup_{T\rightarrow\infty}\frac{1}{T}\mathbb{E}\left[r_t(T)\right]
\end{align}
to find the optimal solution from time $t$ onwards (cost-to-go). The above solution depends only on future energy arrivals {\it after} time $t$, which are, by the memoryless property of the exponential distribution, independent of the events in $[t-\tau,t)$. Only the age and the battery state at time $t$ are needed to solve this problem. This concludes the proof.
\end{Proof}

By Theorem~\ref{thm_inc_rnwl}, focusing on state $(k,0)$ for some $k\leq B-1$ and defining the epochs with respect to this state, problem (\ref{opt_main_inc}) reduces to an optimization over a single epoch. Based on Corollary~\ref{thm_inc_indp} (and Lemma~\ref{thm_state_dep_policy}), we introduce the following notation, which is slightly different than that used in Section~\ref{sec_rbr}.

Once the system goes into state $(k,0)$, for $1\leq k\leq B-1$, at some time $l$, the sensor schedules its next update after $x_k$ time. Since $x_k$ does not depend on the history before time $l$, and cannot depend on the future energy arrivals by the energy causality constraint, we conclude that it is a constant. Now if the first energy arrival in that epoch occurs at time $l+\tau_1$ with $\tau_1>x_k$, the sensor transmits the update at $l+x_k$, whence the state becomes $(k-1,0)$, and if $k\geq2$ the sensor schedules its next update after $x_{k-1}$ time, i.e., at $l+x_k+x_{k-1}$. On the other hand, if the first energy unit arrives relatively early, i.e., $\tau_1\leq x_k$, the state becomes $(k+1,\tau_1)$ at $l+\tau_1$, and the sensor {\it reschedules} the update to be at $l+y_{k+1}(\tau_1)$ instead of $l+x_k$. Note that $y_{k+1}$ only depends on $\tau_1$, since it does not depend on the history before time $l$. If the second energy arrival in that epoch occurs at time $l+\tau_1+\tau_2$ with $\tau_2>y_{k+1}(\tau_1)$, the sensor transmits the update at $l+y_{k+1}(\tau_1)$, whence the state returns to $(k,0)$. On the other hand, if the second energy arrival occurs relatively early as well, i.e., $\tau_2\leq y_{k+1}(\tau_1)$, and if $k\leq B-2$, the state becomes $(k+2,\tau_1+\tau_2)$ at $l+\tau_1+\tau_2$, and the sensor reschedules the update at $l+y_{k+2}(\tau_1+\tau_2)$ instead of $l+y_{k+1}(\tau_1)$.

In summary, the optimal update policy is completely characterized by $B-1$ constants: $\{x_1,x_2,\dots,x_{B-1}\}$, and $B$ functions: $\{y_1(\cdot),y_2(\cdot),\dots,y_B(\cdot)\}$, where $x_k$ represents the scheduled update time after entering state $(k,0)$, and $y_k(t)$ represents the scheduled update time after entering state $(k,t)$ at some time $t$. We emphasize the fact that by Corollary~\ref{thm_inc_indp}, the constants $\{x_k\}$ neither depend on each other, nor on the functions $\{y_k(\cdot)\}$.

\subsection{Renewal State Analysis}

To analyze the optimal solution of our problem, in view of Theorem~\ref{thm_inc_rnwl}, we now need to choose some renewal state $(k,0)$, $k\leq B-1$, and define the epoch with respect to that state. Unlike the random battery recharges problem in Section~\ref{sec_rbr}, unfortunately, there is no choice of $k$ that guarantees a finite number of updates in an epoch; for all choices of $k\leq B-1$ there can possibly be an infinite number of updates in a single epoch. In the sequel, we continue our analysis with state $(0,0)$ as the renewal state and define the epochs with respect to it, i.e., an epoch from now onwards denotes the time between two consecutive visits to state $(0,0)$. We note, however, that any other renewal state choice yields the same results with equivalent complexity. We use the notation $R\left({\bm x},{\bm y}\right)$ and $L\left({\bm x},{\bm y}\right)$ to denote the area under the age curve in a given epoch and its length, respectively, as a function of the constants ${\bm x}\triangleq[x_1,x_2,\dots,x_{B-1}]$ and the functions ${\bm y}\triangleq[y_1,y_2,\dots,y_B]$. Using the strong law of large numbers of renewal processes (renewal reward theorem) \cite[Theorem~3.6.1]{ross_stochastic}, problem (\ref{opt_main_inc}) now reduces to:
\begin{align} \label{opt_rnwl_inc}
\rho=\min_{{\bm x},{\bm y}}\quad&\frac{\mathbb{E}\left[R\left({\bm x},{\bm y}\right)\right]}{\mathbb{E}\left[L\left({\bm x},{\bm y}\right)\right]} \nonumber \\
\mbox{s.t.}\quad &x_k\geq0,\quad1\leq k\leq B-1 \nonumber \\
&y_k(\tau)\geq\tau,\quad1\leq k\leq B.
\end{align}
As in the previous section we introduce the auxiliary parameterized problem:
\begin{align} \label{opt_lmda_inc}
p_B^{\text{ibr}}(\lambda)\triangleq\min_{{\bm x},{\bm y}}\quad&\mathbb{E}\left[R\left({\bm x},{\bm y}\right)\right] -\lambda\mathbb{E}\left[L\left({\bm x},{\bm y}\right)\right] \nonumber \\
\mbox{s.t.}\quad &\text{constraints of (\ref{opt_rnwl_inc})}.
\end{align}
In view of Lemma~\ref{thm_lmda}, we solve for the unique $\lambda^*$ such that $p_B^{\text{ibr}}(\lambda^*)=0$.

One main goal now is to express $\mathbb{E}\left[R\left({\bm x},{\bm y}\right)\right]$ and $\mathbb{E}\left[L\left({\bm x},{\bm y}\right)\right]$ explicitly in terms of ${\bm x}$ and ${\bm y}$ in order to proceed with the optimization. In our previous work \cite{arafa-age-sgl}, we do so for the case $B=2$ through some involved analysis. We note, however, that the analysis approach in \cite{arafa-age-sgl} does not directly extend for general $B$ as it is of a complex combinatorial nature. In what follows, we introduce a novel technique that expresses the objective function of problem (\ref{opt_lmda_inc}) explicitly in terms of ${\bm x}$ and ${\bm y}$ for general $B$, and in fact shortens the analysis in \cite{arafa-age-sgl} for $B=2$.

\begin{figure}
\center
\includegraphics[scale=1]{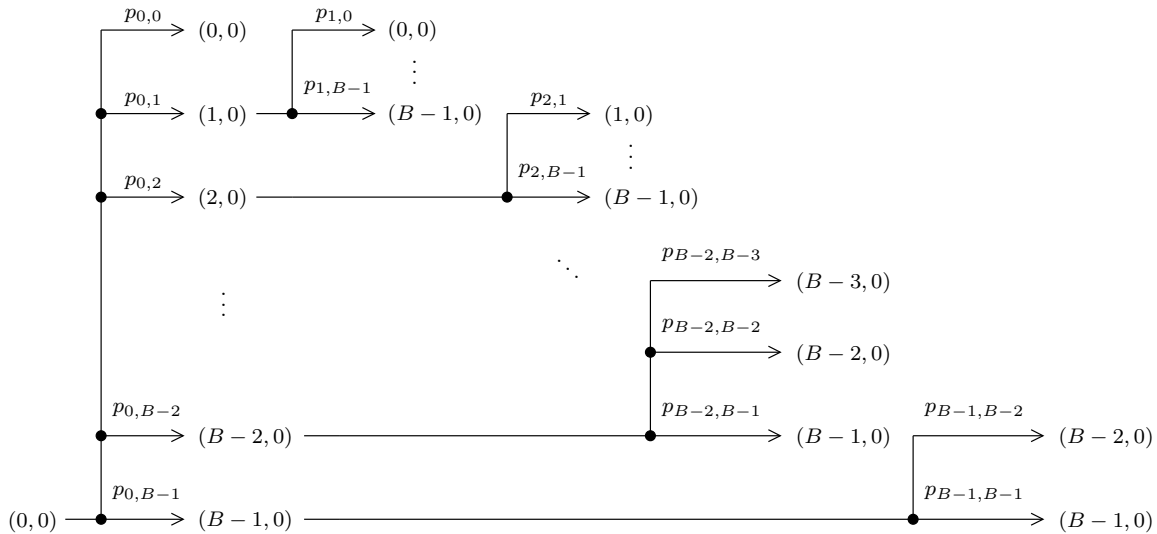}
\caption{Transitions among system states after only {\it one} update. Each transition from state $(i,0)$ to $(j,0)$ occurs with probability $p_{i,j}$ as indicated on the tree branches.}
\label{fig_inc_tree}
\end{figure}

For convenience, we remove the dependency on $\{{\bm x},{\bm y}\}$ in the sequel. Observe that starting from state $(0,0)$ the system can go to any other state $(j,0)$, $0\leq j\leq B-1$, by the next status update, i.e., after only {\it one} update, each with some probability. Then, from state $(j,0)$, $1\leq j\leq B-1$, the system can only go to one of the following states by the next update: $\{(j-1,0),~(j,0),~\dots,(B-1,0)\}$, each with some probability. We denote by $p_{i,j}$ the probability of going from state $(i,0)$ to state $(j,0)$ after one update. Clearly $p_{i,j}=0$ for $j\leq i-2$. We also denote by $r_{i,j}$ and $\ell_{i,j}$ the area under the age curve and the time taken when the system goes from state $(i,0)$ to state $(j,0)$ in one update, respectively. Finally, since the goal is to compute the area under the age curve in an epoch together with the epoch length, we define $R_j$ and $L_j$ as the area under the age curve and the time taken to go from state $(j,0)$ back to $(0,0)$ again (in however many number of updates). See Fig.~\ref{fig_inc_tree} where we depict the relationships between the previous variables/notation in the form of a tree graph. The graph basically represents the transitions between different system states (nodes on the graph) after only {\it one} update, which occur with probabilities indicated on the arrows in the graph that connect the nodes. We emphasize that, for instance, state $(0,0)$ in the first column of the graph is no different than state $(0,0)$ in the second column, and that the arrow connecting them merely represents a loop connecting a state to itself; we chose to expand such loop horizontally for clarity of presentation. From the graph, one can write the following equations:
\begin{align}
\mathbb{E}\left[R\right]&=p_{0,0}\mathbb{E}\left[r_{0,0}\right]+ \sum_{j=1}^{B-1}p_{0,j}\left(\mathbb{E}\left[r_{0,j}\right]+\mathbb{E}\left[R_j\right]\right), \label{eq_inc_R} \\
\mathbb{E}\left[L\right]&=p_{0,0}\mathbb{E}\left[\ell_{0,0}\right]+ \sum_{j=1}^{B-1}p_{0,j}\left(\mathbb{E}\left[\ell_{0,j}\right]+\mathbb{E}\left[L_j\right]\right). \label{eq_inc_L}
\end{align}

Next, we evaluate the above equations. We use the following short-hand notation for nested integrals:
\begin{align} \label{eq_short_hand}
\int\limits_{[a_1,a_2,\dots,a_n]}d{\bm \tau}_1^n \triangleq \int\limits_{\tau_1=0}^{a_1}\int\limits_{\tau_2=0}^{a_2}\dots\int\limits_{\tau_n=0}^{a_n} d\tau_1d\tau_2\dots d\tau_n.
\end{align}
We first begin by the terms $p_{0,j}$, $\mathbb{E}\left[r_{0,j}\right]$, and $\mathbb{E}\left[\ell_{0,j}\right]$, $0\leq j\leq B-1$, which are directly computable as follows. Without loss of generality, let us assume that we start at state $(0,0)$ at time $0$. To go from state $(0,0)$ to $(0,0)$ after one update means that the sensor receives the first energy arrival in the epoch after time $\tau_1$ and then updates after time $y_1(\tau_1)$. This occurs if and only if the second energy arrival after the start of the epoch arriving at time $\tau_1+\tau_2$ occurs relatively late, i.e., $\tau_2>y_1(\tau_1)-\tau_1$. Note that $\tau_1$ and $\tau_2$ are i.i.d. exponential random variables with parameter $1$. Thus,
\begin{align}
p_{0,0}=\mathbb{P}\left(\tau_2>y_1(\tau_1)-\tau_1\right)=\int_{\tau_1=0}^\infty e^{-y_1(\tau_1)}d\tau_1. \label{eq_inc_p_00}
\end{align}
The area under the age curve and the time taken to go from state $(0,0)$ to state $(0,0)$ after one update are respectively given by the expectation of $\frac{1}{2}y_1(\tau_1)^2$ and $y_1(\tau_1)$ {\it conditioned} on the event $\tau_2>y_1(\tau_1)-\tau_1$. Hence,
\begin{align}
p_{0,0}\mathbb{E}\left[r_{0,0}\right]&=p_{0,0}\mathbb{E}\left[\frac{1}{2}y_1(\tau_1)^2\bigg|\tau_2>y_1(\tau_1)-\tau_1\right]=\int_{\tau_1=0}^\infty \frac{1}{2}y_1(\tau_1)^2e^{-y_1(\tau_1)}d\tau_1, \label{eq_inc_r_00} \\
p_{0,0}\mathbb{E}\left[\ell_{0,0}\right]&=p_{0,0}\mathbb{E}\left[y_1(\tau_1)\big|\tau_2>y_1(\tau_1)-\tau_1\right]=\int_{\tau_1=0}^\infty y_1(\tau_1)e^{-y_1(\tau_1)}d\tau_1. \label{eq_inc_l_00}
\end{align}
Next, to go from state $(0,0)$ to $(n,0)$, $1\leq n\leq B-2$, after one update means that the sensor receives $n+1$ energy units consecutively before updating. This occurs if and only if each of the $n+1$ energy units arrive relatively early. That is, after the first arrival at time $\tau_1$ the sensor receives the second arrival at $\tau_1+\tau_2$ with $\tau_2\leq y_1(\tau_1)-\tau_1$, and then the third energy arrival occurs at $\tau_1+\tau_2+\tau_3$ with $\tau_3\leq y_2(\tau_1+\tau_2)-(\tau_1+\tau_2)$, and so on. Only the $(n+2)$th arrival occurs relatively late so that the sensor updates exactly after $n+1$ arrivals, i.e., $\tau_{n+2}>y_{n+1}(\tau_1+\dots+\tau_{n+1})-(\tau_1+\dots+\tau_{n+1})$. Thus, for $1\leq n\leq B-2$
\begin{align}
p_{0,n}&=\mathbb{P}\left(\tau_2\leq y_1(\tau_1)-\tau_1,\tau_3\leq y_2(\tau_1+\tau_2)-(\tau_1+\tau_2),\dots, \right. \nonumber \\
&\hspace{-.25in}\left.\tau_{n+1}\leq y_n(\tau_1+\dots+\tau_n)-(\tau_1+\dots+\tau_n),\tau_{n+2}>y_{n+1}(\tau_1+\dots+\tau_{n+1})-(\tau_1+\dots+\tau_{n+1})\right) \nonumber \\
&=\int_{\tau_1=0}^\infty\int_{\tau_2=0}^{y_1(\tau_1)-\tau_1}\dots\int_{\tau_{n+1}=0}^{y_n(\tau_1+\dots+\tau_n)-(\tau_1+\dots+\tau_n)}e^{-y_{n+1}(\tau_1+\dots+\tau_{n+1})}d\tau_1d\tau_2\dots d\tau_{n+1} \nonumber \\
&=\int\limits_{[\infty,~y_1(\tau_1)-\tau_1,~\dots,~y_n(\tau_1+\dots+\tau_n)-(\tau_1+\dots+\tau_n)]}\hspace{-1in}e^{-y_{n+1}(\tau_1+\dots+\tau_{n+1})}d{\bm \tau}_1^{n+1}, \label{eq_inc_p_0n}
\end{align}
where the last equality is according to the short-hand notation defined in (\ref{eq_short_hand}). Similarly, we have
\begin{align}
p_{0,n}\mathbb{E}\left[r_{0,n}\right]&=\int\limits_{[\infty,~y_1(\tau_1)-\tau_1,~\dots,~y_n(\tau_1+\dots+\tau_n)-(\tau_1+\dots+\tau_n)]}\hspace{-1in}\frac{1}{2}y_{n+1}(\tau_1+\dots+\tau_{n+1})^2e^{-y_{n+1}(\tau_1+\dots+\tau_{n+1})}d{\bm \tau}_1^{n+1}, \label{eq_inc_r_0n} \\
p_{0,n}\mathbb{E}\left[\ell_{0,n}\right]&=\int\limits_{[\infty,~y_1(\tau_1)-\tau_1,~\dots,~y_n(\tau_1+\dots+\tau_n)-(\tau_1+\dots+\tau_n)]}\hspace{-1in}y_{n+1}(\tau_1+\dots+\tau_{n+1})e^{-y_{n+1}(\tau_1+\dots+\tau_{n+1})}d{\bm \tau}_1^{n+1}. \label{eq_inc_l_0n}
\end{align}
Finally, to go from state $(0,0)$ to $(B-1,0)$ after one update means that the sensor receives $B$ consecutive energy units, i.e., until its battery is full, with relatively early inter-arrival times. Thus,
\begin{align}
p_{0,B-1}&=\mathbb{P}\left(\tau_2\leq y_1(\tau_1)-\tau_1,\tau_3\leq y_2(\tau_1+\tau_2)-(\tau_1+\tau_2),\dots, \right. \nonumber \\
&\hspace{2in}\left.\tau_{B}\leq y_{B-1}(\tau_1+\dots+\tau_{B-1})-(\tau_1+\dots+\tau_{B-1})\right) \nonumber \\
&=\int\limits_{[\infty,~y_1(\tau_1)-\tau_1,~\dots,~y_{B-1}(\tau_1+\dots+\tau_{B-1})-(\tau_1+\dots+\tau_{B-1})]}\hspace{-1in}e^{-(\tau_1+\dots+\tau_{B})}d{\bm \tau}_1^B. \label{eq_inc_p_0B1}
\end{align}
Similarly, we have
\begin{align}
p_{0,B-1}\mathbb{E}\left[r_{0,B-1}\right]&=\int\limits_{[\infty,~y_1(\tau_1)-\tau_1,~\dots,~y_{B-1}(\tau_1+\dots+\tau_{B-1})-(\tau_1+\dots+\tau_{B-1})]}\hspace{-1in}\frac{1}{2}y_B(\tau_1+\dots+\tau_B)^2e^{-(\tau_1+\dots+\tau_{B})}d{\bm \tau}_1^B, \label{eq_inc_r_0B1} \\
p_{0,B-1}\mathbb{E}\left[l_{0,B-1}\right]&=\int\limits_{[\infty,~y_1(\tau_1)-\tau_1,~\dots,~y_{B-1}(\tau_1+\dots+\tau_{B-1})-(\tau_1+\dots+\tau_{B-1})]}\hspace{-1in}y_B(\tau_1+\dots+\tau_B)e^{-(\tau_1+\dots+\tau_{B})}d{\bm \tau}_1^B. \label{eq_inc_l_0B1}
\end{align}

We now move on to computing the terms $\mathbb{E}\left[R_j\right]$ and $\mathbb{E}\left[L_j\right]$, $1\leq j\leq B-1$. These are not as directly computable as the terms $p_{0,j}$, $\mathbb{E}\left[r_{0,j}\right]$, and $\mathbb{E}\left[l_{0,j}\right]$, $0\leq j\leq B-1$, and are evaluated via recursive formulas from the tree in Fig.~\ref{fig_inc_tree}, which we explain as follows. We notice that the tree starts with one root node, state $(0,0)$, and that it has all other possible states in its second stage. Starting from that second stage, and focusing on the terms $\mathbb{E}\left[R_j\right]$, $1\leq j\leq B-1$, for now (calculations for the terms $\mathbb{E}\left[L_j\right]$, $1\leq j\leq B-1$ are analogous), one can write
\begin{align}
\mathbb{E}\left[R_1\right]&=\sum_{i=0}^{B-1}p_{1,i}\mathbb{E}\left[r_{1,i}\right] + \sum_{i=1}^{B-1}p_{1,i}\mathbb{E}\left[R_i\right], \label{eq_inc_R1_rec} \\
\mathbb{E}\left[R_j\right]&=\sum_{i=j-1}^{B-1}p_{j,i}\mathbb{E}\left[r_{j,i}\right] + \sum_{i=j-1}^{B-1}p_{j,i}\mathbb{E}\left[R_i\right],\quad2\leq j\leq B-1. \label{eq_inc_Rj_rec}
\end{align}
Next, we begin from the last equation, i.e., (\ref{eq_inc_Rj_rec}) with $j=B-1$, and make use of the fact that $p_{B-1,B-2}+p_{B-1,B-1}=1$ to write
\begin{align} \label{eq_inc_rec_b1}
\mathbb{E}\left[R_{B-1}\right]=\mathbb{E}\left[r_{B-1,B-2}\right]+\frac{1}{p_{B-1,B-2}}p_{B-1,B-1}\mathbb{E}\left[r_{B-1,B-1}\right]+\mathbb{E}\left[R_{B-2}\right].
\end{align}
We then work our way backwards; we substitute (\ref{eq_inc_rec_b1}) in (\ref{eq_inc_Rj_rec}) with $j=B-2$, and again make use of the fact that $p_{B-2,B-3}+p_{B-2,B-2}+p_{B-2,B-1}=1$, to get after some simple manipulations that
\begin{align} \label{eq_inc_rec_b2}
\mathbb{E}\left[R_{B-2}\right]=&\mathbb{E}\left[r_{B-2,B-3}\right]+\frac{1}{p_{B-2,B-3}}p_{B-2,B-2}\mathbb{E}\left[r_{B-2,B-2}\right]+\frac{1}{p_{B-2,B-3}}p_{B-2,B-1}\mathbb{E}\left[r_{B-2,B-1}\right] \nonumber \\
&+\frac{p_{B-2,B-1}}{p_{B-2,B-3}}\left(\mathbb{E}\left[r_{B-1,B-2}\right]+\frac{1}{p_{B-1,B-2}}p_{B-1,B-1}\mathbb{E}\left[r_{B-1,B-1}\right]\right)+\mathbb{E}\left[R_{B-3}\right].
\end{align}
We then substitute (\ref{eq_inc_rec_b1}) and (\ref{eq_inc_rec_b2}) in (\ref{eq_inc_Rj_rec}) with $j=B-3$ to get $\mathbb{E}\left[R_{B-3}\right]$ in terms of $\mathbb{E}\left[R_{B-4}\right]$, and so on. Continuing this way recursively, we get $B-2$ equations with each equation having a term $\mathbb{E}\left[R_j\right]$ in terms of $\mathbb{E}\left[R_{j-1}\right]$, $2\leq j\leq B-1$, which can be written as follows:
\begin{align} \label{eq_inc_Rj_Rjbar}
\mathbb{E}\left[R_j\right]=\bar{R}_j+\mathbb{E}\left[R_{j-1}\right], \quad 2\leq j\leq B-1,
\end{align}
with $\bar{R}_j$ defined as
\begin{align} \label{eq_inc_R_bar}
\bar{R}_j\triangleq&\mathbb{E}\left[r_{j,j-1}\right]+\frac{1}{p_{j,j-1}}\sum_{i=j}^{B-1}p_{j,i}\mathbb{E}\left[r_{j,i}\right] \nonumber \\
&+c_{j,j+1}\left(\mathbb{E}\left[r_{j+1,j}\right]+\frac{1}{p_{j+1,j}}\sum_{i=j+1}^{B-1}p_{j+1,i}\mathbb{E}\left[r_{j+1,i}\right]\right) \nonumber \\
&+c_{j,j+2}\left(\mathbb{E}\left[r_{j+2,j+1}\right]+\frac{1}{p_{j+2,j+1}}\sum_{i=j+2}^{B-1}p_{j+2,i}\mathbb{E}\left[r_{j+2,i}\right]\right) \nonumber \\
&+\dots \nonumber \\
&+c_{j,B-1}\left(\mathbb{E}\left[r_{B-1,B-2}\right]+\frac{1}{p_{B-1,B-2}}p_{B-1,B-1}\mathbb{E}\left[r_{B-1,B-1}\right]\right),\quad 2\leq j\leq B-1,
\end{align}
and with the constants $c_{n,m}$ defined as
\begin{align}
c_{n,n+1}&\triangleq\frac{1}{p_{n,n-1}}\sum_{i=n+1}^{B-1}p_{n,i}, \label{eq_inc_c_n_n1} \\
c_{n,m}&\triangleq\sum_{i\in\mathcal{P}\left({\{n+1,\dots,m-1\}}\right)}\frac{1}{\prod\limits_{\substack{j=n \\ j\notin i}}^mp_{j,j-1}}\prod_{\substack{j=n \\ j\notin i}}^{m-1}~\sum_{l=j_i(m)}^{B-1}p_{j,l},\quad n+2\leq m\leq B-1, \label{eq_inc_c_n_m}
\end{align}
where $\mathcal{P}\left({\omega}\right)$ is the power set of the set $\omega$ (note that the summand $i$ in (\ref{eq_inc_c_n_m}) is actually a subset), and $j_i(m)\triangleq\min\left\{\{j+1,\dots,m\}\setminus i\right\}$. Observe that one can rewrite the equations in (\ref{eq_inc_Rj_Rjbar}) slightly differently after some simple backward substitutions as follows:
\begin{align} \label{eq_inc_Rj_R1}
\mathbb{E}\left[R_j\right]=\bar{R}_j+\bar{R}_{j-1}+\dots+\bar{R}_2+\mathbb{E}\left[R_1\right],\quad 2\leq j\leq B-1.
\end{align}
Therefore, what remains to evaluate the terms $\mathbb{E}\left[R_j\right]$, $2\leq j\leq B-1$, is to evaluate $\mathbb{E}\left[R_1\right]$. We do so by substituting all $B-2$ equations of (\ref{eq_inc_Rj_R1}) back in (\ref{eq_inc_R1_rec}) to finally get
\begin{align} \label{eq_inc_R1}
\mathbb{E}\left[R_1\right]=&\mathbb{E}\left[r_{1,0}\right]+\frac{1}{p_{1,0}}\sum_{j=1}^{B-1}p_{1,j}\mathbb{E}\left[r_{1,j}\right] \nonumber \\
&+c_{12}\left(\mathbb{E}\left[r_{2,1}\right]+\frac{1}{p_{21}}\sum_{j=2}^{B-1}p_{2,j}\mathbb{E}\left[r_{2,j}\right]\right) \nonumber \\
&+c_{1,3}\left(\mathbb{E}\left[r_{3,1}\right]+\frac{1}{p_{3,1}}\sum_{j=3}^{B-1}p_{3,j}\mathbb{E}\left[r_{3,j}\right]\right) \nonumber \\
&+\dots \nonumber \\
&+c_{1,B-1}\left(\mathbb{E}\left[r_{B-1,B-2}\right]+\frac{1}{p_{B-1,B-2}}p_{B-1,B-1}\mathbb{E}\left[r_{B-1,B-1}\right]\right),
\end{align}
where the constants $c_{1,m}$, $2\leq m\leq B-1$, are as defined in (\ref{eq_inc_c_n_n1}) and (\ref{eq_inc_c_n_m}) for $n=1$. Equations (\ref{eq_inc_Rj_R1}) and (\ref{eq_inc_R1}) fully characterize the terms $\mathbb{E}\left[R_j\right]$, $1\leq j\leq B-1$. As for the terms $\mathbb{E}\left[L_j\right]$, $1\leq j\leq B-1$, they can be completely characterized in the exact same recursive manner as above with only switching the terms $r_{j,i}$ by $\ell_{j,i}$ and defining $\bar{L}_j$ analogously to $\bar{R}_j$ and so on.

Using (\ref{eq_inc_Rj_R1}) and (\ref{eq_inc_R1}) in (\ref{eq_inc_R}), we get
\begin{align}
\mathbb{E}\left[R\right]=&\sum_{j=0}^{B-1}p_{0,j}\mathbb{E}\left[r_{0,j}\right] \nonumber \\
&+\left(\sum_{j=1}^{B-1}p_{0,j}\right)\left(\mathbb{E}\left[r_{1,0}\right]+\frac{1}{p_{1,0}}\sum_{j=1}^{B-1}p_{1,j}\mathbb{E}\left[r_{1,j}\right]\right) \nonumber \\
&+\left(\sum_{j=2}^{B-1}p_{0,j}+c_{1,2}\sum_{j=1}^{B-1}p_{0,j}\right)\left(\mathbb{E}\left[r_{21}\right]+\frac{1}{p_{21}}\sum_{j=2}^{B-1}p_{2,j}\mathbb{E}\left[r_{2,j}\right]\right) \nonumber \\
&+\dots \nonumber \\
&+\left(\sum_{j=n}^{B-1}p_{0,j}+c_{n-1,n}\sum_{j=n-1}^{B-1}p_{0,j}+\dots+c_{1,n}\sum_{j=1}^{B-1}p_{0,j}\right) \left(\mathbb{E}\left[r_{n,n-1}\right]+\frac{1}{p_{n,n-1}}\sum_{j=n}^{B-1}p_{n,j}\mathbb{E}\left[r_{n,j}\right]\right) \nonumber \\
&+\dots \nonumber \\
&+\left(\sum_{j=n}^{B-1}p_{0,B-1}+c_{B-2,B-1}\left(p_{0,B-2}+p_{0,B-1}\right)+\dots+c_{1,B-1}\sum_{j=1}^{B-1}p_{0,j}\right) \nonumber \\
&\hspace{1.75in}\times \left(\mathbb{E}\left[r_{B-1,B-2}\right]+\frac{1}{p_{B-1,B-2}}p_{B-1,B-1}\mathbb{E}\left[r_{B-1,B-1}\right]\right). \label{eq_inc_R_final}
\end{align}
Similarly, we also have
\begin{align}
\mathbb{E}\left[L\right]=&\sum_{j=0}^{B-1}p_{0,j}\mathbb{E}\left[\ell_{0,j}\right] \nonumber \\
&+\left(\sum_{j=1}^{B-1}p_{0,j}\right)\left(\mathbb{E}\left[\ell_{1,0}\right]+\frac{1}{p_{1,0}}\sum_{j=1}^{B-1}p_{1,j}\mathbb{E}\left[\ell_{1,j}\right]\right) \nonumber \\
&+\left(\sum_{j=2}^{B-1}p_{0,j}+c_{1,2}\sum_{j=1}^{B-1}p_{0,j}\right)\left(\mathbb{E}\left[\ell_{21}\right]+\frac{1}{p_{21}}\sum_{j=2}^{B-1}p_{2,j}\mathbb{E}\left[\ell_{2,j}\right]\right) \nonumber \\
&+\dots \nonumber \\
&+\left(\sum_{j=n}^{B-1}p_{0,j}+c_{n-1,n}\sum_{j=n-1}^{B-1}p_{0,j}+\dots+c_{1,n}\sum_{j=1}^{B-1}p_{0,j}\right) \left(\mathbb{E}\left[\ell_{n,n-1}\right]+\frac{1}{p_{n,n-1}}\sum_{j=n}^{B-1}p_{n,j}\mathbb{E}\left[\ell_{n,j}\right]\right) \nonumber \\
&+\dots \nonumber \\
&+\left(\sum_{j=n}^{B-1}p_{0,B-1}+c_{B-2,B-1}\left(p_{0,B-2}+p_{0,B-1}\right)+\dots+c_{1,B-1}\sum_{j=1}^{B-1}p_{0,j}\right) \nonumber \\
&\hspace{1.75in}\times \left(\mathbb{E}\left[\ell_{B-1,B-2}\right]+\frac{1}{p_{B-1,B-2}}p_{B-1,B-1}\mathbb{E}\left[\ell_{B-1,B-1}\right]\right). \label{eq_inc_L_final}
\end{align}

What remains now is to characterize the terms $p_{j,n}$, $\mathbb{E}\left[r_{j,n}\right]$, and $\mathbb{E}\left[\ell_{j,n}\right]$, $1\leq n\leq B-1$, $1\leq j\leq B-1$. These are directly computable terms via the same arguments involved before in the computations of the terms $p_{0,j}$, $\mathbb{E}\left[r_{0,j}\right]$, and $\mathbb{E}\left[\ell_{0,j}\right]$, $0\leq j\leq B-1$. We first consider the special case when the system goes from state $(j,0)$ to state $(j-1,0)$, $1\leq j\leq B-1$, after one update. This occurs if and only if the first energy arrival arriving $\tau_1$ time units after going through state $(j,0)$ occurs relatively late, i.e., the sensor submits an update after $x_j$ time units {\it before} receiving such energy unit. Since $\tau_1$ is an exponential random variable with parameter $1$, we have
\begin{align}
p_{j,j-1}=\mathbb{P}\left(\tau_1>x_j\right)=e^{-x_j}. \label{eq_inc_p_jj1}
\end{align}
The area under age curve and the time taken to go from state $(j,0)$ to state $(j-1,0)$, $1\leq j\leq B-1$, after one update are respectively given by the expectation of the constants $\frac{1}{2}x_j^2$ and $x_j$ {\it conditioned} on the event $\tau_1>x_j$. Hence,
\begin{align}
\mathbb{E}\left[r_{j,j-1}\right]&=\mathbb{E}\left[\frac{1}{2}x_j^2\bigg|\tau_1>x_j\right] =\frac{1}{2}x_j^2, \label{eq_inc_r_jj1} \\
\mathbb{E}\left[\ell_{j,j-1}\right]&=\mathbb{E}\left[x_j\big|\tau_1>x_j\right] =x_j. \label{eq_inc_l_jj1}
\end{align}
Next, we consider the case of going from state $(j,0)$ to $(n,0)$, $1\leq j\leq B-1$, $j\leq n\leq B-2$, after one update. We proceed similar to the way we derived the terms $p_{0,n}$, $\mathbb{E}\left[r_{0,n}\right]$, and $\mathbb{E}\left[\ell_{0,n}\right]$ in (\ref{eq_inc_p_00}), (\ref{eq_inc_r_00}), and (\ref{eq_inc_l_00}), respectively, for $j=n=0$; in (\ref{eq_inc_p_0n}), (\ref{eq_inc_r_0n}), and (\ref{eq_inc_l_0n}), respectively, for $j=0$ and $1\leq n\leq B-1$; and in (\ref{eq_inc_p_0B1}), (\ref{eq_inc_r_0B1}), and (\ref{eq_inc_l_0B1}), respectively, for $j=0$ and $n=B-1$. We state the results in what follows. First, for $1\leq j\leq B-2$ and $n=j$, we have
\begin{align}
p_{j,j}&=\mathbb{P}\left(\tau_1\leq x_j,\tau_2> y_{j+1}(\tau_1)-\tau_1\right)=\int_{\tau_1=0}^{x_j}e^{-y_{j+1}(\tau_1)}d\tau_1, \label{eq_inc_p_jj} \\
p_{j,j}\mathbb{E}\left[r_{j,j}\right]&=\int_{\tau_1=0}^{x_j}\frac{1}{2}y_{j+1}(\tau_1)^2e^{-y_{j+1}(\tau_1)}d\tau_1, \label{eq_inc_r_jj} \\
p_{j,j}\mathbb{E}\left[\ell_{j,j}\right]&=\int_{\tau_1=0}^{x_j}y_{j+1}(\tau_1)e^{-y_{j+1}(\tau_1)}d\tau_1. \label{eq_inc_l_jj}
\end{align}
Next, for $1\leq j\leq B-2$ and $j+1\leq n\leq B-2$, we have
\begin{align}
p_{j,n}&=\mathbb{P}\left(\tau_1\leq x_j,\tau_2\leq y_{j+1}(\tau_1)-\tau_1,\tau_3\leq y_{j+2}(\tau_1+\tau_2)-(\tau_1+\tau_2),\dots, \right. \nonumber \\
&\hspace{1in}\left.\tau_{n-j+1}\leq y_n(\tau_1+\dots+\tau_{n-j})-(\tau_1+\dots+\tau_{n-j}), \right. \nonumber \\
&\hspace{1in}\left. \tau_{n-j+2}>y_{n+1}(\tau_1+\dots+\tau_{n-j+1})-(\tau_1+\dots+\tau_{n-j+1})\right) \nonumber \\
&=\int\limits_{[x_j,~y_{j+1}(\tau_1)-\tau_1,~\dots,~y_n(\tau_1+\dots+\tau_{n-j})-(\tau_1+\dots+\tau_{n-j})]}\hspace{-1in}e^{-y_{n+1}(\tau_1+\dots+\tau_{n-j+1})}d{\bm \tau}_1^{n-j+1}, \label{eq_inc_p_jn} \\
p_{j,n}\mathbb{E}\left[r_{j,n}\right]&=\int\limits_{[x_j,~y_{j+1}(\tau_1)-\tau_1,~\dots,~y_n(\tau_1+\dots+\tau_{n-j})-(\tau_1+\dots+\tau_{n-j})]}\hspace{-1in}\frac{1}{2}y_{n+1}(\tau_1+\dots+\tau_{n-j+1})^2e^{-y_{n+1}(\tau_1+\dots+\tau_{n-j+1})}d{\bm \tau}_1^{n-j+1}, \label{eq_inc_r_jn} \\
p_{j,n}\mathbb{E}\left[\ell_{j,n}\right]&=\int\limits_{[x_j,~y_{j+1}(\tau_1)-\tau_1,~\dots,~y_n(\tau_1+\dots+\tau_{n-j})-(\tau_1+\dots+\tau_{n-j})]}\hspace{-1in}y_{n+1}(\tau_1+\dots+\tau_{n-j+1})e^{-y_{n+1}(\tau_1+\dots+\tau_{n-j+1})}d{\bm \tau}_1^{n-j+1}. \label{eq_inc_l_jn}
\end{align}
Then, for $1\leq j\leq B-2$ and $n=B-1$, we have
\begin{align}
p_{j,B-1}&=\mathbb{P}\left(\tau_1\leq x_j,\tau_2\leq y_{j+1}(\tau_1)-\tau_1,\tau_3\leq y_{j+2}(\tau_1+\tau_2)-(\tau_1+\tau_2),\dots, \right. \nonumber \\
&\hspace{1in}\left.\tau_{B-j}\leq y_{B-1}(\tau_1+\dots+\tau_{B-1-j})-(\tau_1+\dots+\tau_{B-1-j})\right) \nonumber \\
&=\int\limits_{[x_j,~y_{j+1}(\tau_1)-\tau_1,~\dots,~y_{B-1}(\tau_1+\dots+\tau_{B-1-j})-(\tau_1+\dots+\tau_{B-1-j})]}\hspace{-1in}e^{-(\tau_1+\dots+\tau_{B-j})}d{\bm \tau}_1^{B-j}, \label{eq_inc_p_jB1} \\
p_{j,B-1}\mathbb{E}\left[r_{j,B-1}\right]&=\int\limits_{[x_j,~y_{j+1}(\tau_1)-\tau_1,~\dots,~y_{B-1}(\tau_1+\dots+\tau_{B-1-j})-(\tau_1+\dots+\tau_{B-1-j})]}\hspace{-1in}\frac{1}{2}y_B(\tau_1+\dots+\tau_{B-j})^2e^{-(\tau_1+\dots+\tau_{B-j})}d{\bm \tau_1}^{B-j}, \label{eq_inc_r_jB1} \\
p_{j,B-1}\mathbb{E}\left[\ell_{j,B-1}\right]&=\int\limits_{[x_j,~y_{j+1}(\tau_1)-\tau_1,~\dots,~y_{B-1}(\tau_1+\dots+\tau_{B-1-j})-(\tau_1+\dots+\tau_{B-1-j})]}\hspace{-1in}y_B(\tau_1+\dots+\tau_{B-j})e^{-(\tau_1+\dots+\tau_{B-j})}d{\bm \tau}_1^{B-j}. \label{eq_inc_l_jB1}
\end{align}
Finally, for $j=n=B-1$, we have
\begin{align}
p_{B-1,B-1}\mathbb{E}\left[r_{B-1,B-1}\right]&=\int_{\tau_1=0}^{x_{B-1}}\frac{1}{2}y_B(\tau_1)^2e^{-\tau_1}d\tau_1, \label{eq_inc_r_B1B1} \\
p_{B-1,B-1}\mathbb{E}\left[\ell_{B-1,B-1}\right]&=\int_{\tau_1=0}^{x_{B-1}}y_B(\tau_1)e^{-\tau_1}d\tau_1. \label{eq_inc_l_B1B1}
\end{align}
Observe that the term $p_{B-1,B-1}$ does not appear individually in (\ref{eq_inc_R_final}) or (\ref{eq_inc_L_final}).

We now have every term needed to fully characterize the objective function of problem (\ref{opt_lmda_inc}) in terms of the constants ${\bm x}$ and the functions ${\bm y}$. We do so by basically substituting (\ref{eq_inc_p_00})-(\ref{eq_inc_l_0B1}), (\ref{eq_inc_c_n_n1})-(\ref{eq_inc_c_n_m}), and (\ref{eq_inc_p_jj1})-(\ref{eq_inc_l_B1B1}) in (\ref{eq_inc_R_final}) and (\ref{eq_inc_L_final}). In the next subsection, we characterize the optimal constants ${\bm x}$ and functions ${\bm y}$ that solve problem (\ref{opt_lmda_inc}).

\subsection{Threshold Policies}

We introduce the following Lagrangian for problem (\ref{opt_lmda_inc}) \cite{boyd}:
\begin{align} \label{eq_lagrange_inc}
\mathcal{L}=\mathbb{E}\left[R\right]-\lambda\mathbb{E}\left[L\right]-\sum_{i=1}^{B-1}\eta_ix_i-\sum_{i=1}^B\int_{\tau=0}^\infty\gamma_i(\tau)\left(y_i(\tau)-\tau\right)d\tau,
\end{align}
where $\{\eta_i\}$ and $\{\gamma_i(\cdot)\}$ are non-negative Lagrange multipliers. We now proceed by taking derivative of the Lagrangian with respect to each variable and equating it to $0$ in a specific alternating order between the functions ${\bm y}$ and the constants ${\bm x}$. Specifically, we start by taking derivative of the Lagrangian with respect to $y_B(t)$ first, followed by $x_{B-1}$, and then $y_{B-1}(t)$, and then $x_{B-2}$, and so on until $x_1$ and $y_1(t)$. The reason is that, as we explicitly illustrate below, this specific order allows writing each variable only in terms of the preceding variables in the order, which would be already evaluated in terms of $\lambda$.

For simplicity of presentation, we illustrate this methodology by focusing on the case of $B=4$ energy units. This case is sufficiently general in the sense that the techniques invoked in characterizing its optimal solution can be readily extended to any higher value of the battery capacity. For $B=4$, the objective function is given by
\begin{align} \label{eq_inc_b4_obj}
\mathbb{E}\left[R\right]-\lambda\mathbb{E}\left[L\right]=&\sum_{j=0}^3p_{0,j}(\mathbb{E}[r_{0,j}]-\lambda\mathbb{E}[\ell_{0,j}])+\sum_{j=1}^3p_{0,j}\left(\mathbb{E}[R_1]-\lambda\mathbb{E}[L_1]\right) \nonumber \\
&+\sum_{j=2}^3p_{0,j}\left(\bar{R}_2-\lambda\bar{L}_2\right)+p_{0,3}\left(\bar{R}_3-\lambda\bar{L}_3\right).
\end{align}
We now write down the terms constituting the above objective function explicitly in terms of the optimization variables $\{x_1,x_2,x_3\}$ and $\{y_1,y_2,y_3,y_4\}$. We first start by
\begin{align} \label{eq_inc_b4_obj1}
&\sum_{j=0}^3p_{0,j}(\mathbb{E}[r_{0,j}]-\lambda\mathbb{E}[\ell_{0,j}])=\int_{\tau=0}^{\infty}\left(\frac{1}{2}y_1(\tau)^2-\lambda y_1(\tau)\right)e^{-y_1(\tau)}d\tau \nonumber \\
&+\int\limits_{[\infty,y_1(\tau_1)-\tau_1]}\left(\frac{1}{2}y_2(\tau_1+\tau_2)^2-\lambda y_2(\tau_1+\tau_2)\right)e^{-y_2(\tau_1+\tau_2)}d{\bm \tau}_1^2 \nonumber \\
&+\int\limits_{[\infty,y_1(\tau_1)-\tau_1,y_2(\tau_1+\tau_2)-(\tau_1+\tau_2)]} \hspace{-.5in} \left(\frac{1}{2}y_3(\tau_1+\tau_2+\tau_3)^2-\lambda y_3(\tau_1+\tau_2+\tau_3)\right)e^{-y_3(\tau_1+\tau_2+\tau_3)}d{\bm \tau}_1^3 \nonumber \\
&+\int\limits_{[\infty,y_1(\tau_1)-\tau_1,y_2(\tau_1+\tau_2)-(\tau_1+\tau_2),y_3(\tau_1+\tau_2+\tau_3)-(\tau_1+\tau_2+\tau_3)]} \hspace{-1.25in} \left(\frac{1}{2}y_4(\tau_1+\dots+\tau_4)^2-\lambda y_4(\tau_1+\dots+\tau_4)\right)e^{-(\tau_1+\dots+\tau_4)}d{\bm \tau}_1^4.
\end{align}
Then, we have
\begin{align}
p_{0,1}=&\int\limits_{[\infty,y_1(\tau_1)-\tau_1]}e^{-y_2(\tau_1+\tau_2)}d{\bm \tau}_1^2, \\
p_{0,2}=&\int\limits_{[\infty,y_1(\tau_1)-\tau_1,y_2(\tau_1+\tau_2)-(\tau_1+\tau_2)]} \hspace{-.5in} e^{-y_3(\tau_1+\tau_2+\tau_3)}d{\bm \tau}_1^3, \\
p_{0,3}=&\int\limits_{[\infty,y_1(\tau_1)-\tau_1,y_2(\tau_1+\tau_2)-(\tau_1+\tau_2),y_3(\tau_1+\tau_2+\tau_3)-(\tau_1+\tau_2+\tau_3)]} \hspace{-1.25in} e^{-(\tau_1+\dots+\tau_4)}d{\bm \tau}_1^4.
\end{align}
Next, by (\ref{eq_inc_R1}) we have
\begin{align}
\mathbb{E}[R_1]&-\lambda\mathbb{E}[L_1] \nonumber \\
=&\mathbb{E}[r_{1,0}]-\lambda\mathbb{E}[\ell_{1,0}]+\frac{1}{p_{1,0}}\sum_{j=1}^3p_{1,j}\left(\mathbb{E}[r_{1,j}]-\lambda\mathbb{E}[\ell_{1,j}]\right) \nonumber \\
&+\frac{p_{1,2}+p_{1,3}}{p_{1,0}}\left(\mathbb{E}[r_{2,1}]-\lambda\mathbb{E}[\ell_{2,1}]+\frac{1}{p_{2,1}}\sum_{j=2}^3p_{2,j}\left(\mathbb{E}[r_{2,j}]-\lambda\mathbb{E}[\ell_{2,j}]\right)\right) \nonumber \\
&+\left(\frac{(p_{1,2}+p_{1,3})p_{2,3}}{p_{2,1}p_{1,0}}+\frac{p_{1,3}}{p_{1,0}}\right)\left(\mathbb{E}[r_{3,2}]-\lambda\mathbb{E}[\ell_{3,2}]+\frac{1}{p_{3,2}}\left(\mathbb{E}[r_{3,3}]-\lambda\mathbb{E}[\ell_{3,3}]\right)\right) \nonumber \\
=&\frac{1}{2}x_1^2-\lambda x_1+e^{x_1}\Bigg(\int_{\tau=0}^{x_1}\left(\frac{1}{2}y_2(\tau)^2-\lambda y_2(\tau)\right)e^{-y_2(\tau)}d\tau \nonumber \\ 
&\hspace{.25in}+\int\limits_{[x_1,y_2(\tau_1)-\tau_1]}\left(\frac{1}{2}y_3(\tau_1+\tau_2)^2-\lambda y_3(\tau_1+\tau_2)\right)e^{-y_3(\tau_1+\tau_2)}d{\bm \tau}_1^2 \nonumber \\
&\hspace{.25in}+\int\limits_{[x_1,y_2(\tau_1)-\tau_1,y_3(\tau_1+\tau_2)-(\tau_1+\tau_2)]} \hspace{-.5in} \left(\frac{1}{2}y_4(\tau_1+\tau_2+\tau_3)^2-\lambda y_4(\tau_1+\tau_2+\tau_3)\right)e^{-(\tau_1+\tau_2+\tau_3)}d{\bm \tau}_1^3 \Bigg) \nonumber \\
&+(p_{1,2}+p_{1,3})e^{x_1}\Bigg(\frac{1}{2}x_2^2-\lambda x_2+e^{x_2}\bigg(\int_{\tau=0}^{x_2}\left(\frac{1}{2}y_3(\tau)^2-\lambda y_3(\tau)\right)e^{-y_3(\tau)}d\tau \nonumber \\
&\hspace{.25in}+\int\limits_{[x_2,y_3(\tau_1)-\tau_1]}\left(\frac{1}{2}y_4(\tau_1+\tau_2)^2-\lambda y_4(\tau_1+\tau_2)\right)e^{-(\tau_1+\tau_2)}d{\bm \tau}_1^2\bigg)\Bigg) \nonumber \\
&+\left((p_{1,2}+p_{1,3})p_{2,3}e^{x_2}e^{x_1}+p_{1,3}e^{x_1}\right)\left(\frac{1}{2}x_3^2-\lambda x_3+e^{x_3}\int_{\tau=0}^{x_3}\left(\frac{1}{2}y_4(\tau)^2-\lambda y_4(\tau)\right)e^{-\tau}d\tau\right),
\end{align}
where $p_{1,2}$, $p_{1,3}$, and $p_{2,3}$ are given by
\begin{align}
p_{1,2}=&\int\limits_{[x_1,y_2(\tau_1)-\tau_1]}e^{-y_3(\tau_1+\tau_2)}d{\bm \tau}_1^2, \\
p_{1,3}=&\int\limits_{[x_1,y_2(\tau_1)-\tau_1,y_3(\tau_1+\tau_2)-(\tau_1+\tau_2)]} \hspace{-.5in} e^{-(\tau_1+\tau_2+\tau_3)}d{\bm \tau}_1^3, \\
p_{2,3}=&\int\limits_{[x_2,y_3(\tau_1)-\tau_1]}e^{-(\tau_1+\tau_2)}d{\bm \tau}_1^2. 
\end{align}
Finally, by (\ref{eq_inc_R_bar}) we have
\begin{align} 
\bar{R}_2-\lambda\bar{L}_2=&\mathbb{E}[r_{2,1}]-\lambda\mathbb{E}[\ell_{2,1}]+\frac{1}{p_{2,1}}\sum_{j=2}^3p_{2,j}\left(\mathbb{E}[r_{2,j}]-\lambda\mathbb{E}[\ell_{2,j}]\right) \nonumber \\
&+\frac{p_{2,3}}{p_{2,1}}\left(\mathbb{E}[r_{3,2}]-\lambda\mathbb{E}[\ell_{3,2}]+\frac{1}{p_{3,2}}\left(\mathbb{E}[r_{3,3}]-\lambda\mathbb{E}[\ell_{3,3}]\right)\right) \nonumber \\
=&\frac{1}{2}x_2^2-\lambda x_2+e^{x_2}\Bigg(\int_{\tau=0}^{x_2}\left(\frac{1}{2}y_3(\tau)^2-\lambda y_3(\tau)\right)e^{-y_3(\tau)}d\tau \nonumber \\
&\hspace{.25in}+\int\limits_{[x_2,y_3(\tau_1)-\tau_1]}\left(\frac{1}{2}y_4(\tau_1+\tau_2)^2-\lambda y_4(\tau_1+\tau_2)\right)e^{-(\tau_1+\tau_2)}d{\bm \tau}_1^2\Bigg) \nonumber \\
&+p_{2,3}e^{x_2}\left(\frac{1}{2}x_3^2-\lambda x_3+e^{x_3}\int_{\tau=0}^{x_3}\left(\frac{1}{2}y_4(\tau)^2-\lambda y_4(\tau)\right)e^{-\tau}d\tau\right),
\end{align}
and
\begin{align} \label{eq_inc_b4_obj2}
\bar{R}_3-\lambda\bar{L}_3=&\mathbb{E}[r_{3,2}]-\lambda\mathbb{E}[\ell_{3,2}]+\frac{1}{p_{3,2}}\left(\mathbb{E}[r_{3,3}]-\lambda\mathbb{E}[\ell_{3,3}]\right) \nonumber \\
=&\frac{1}{2}x_3^2-\lambda x_3+e^{x_3}\int_{\tau=0}^{x_3}\left(\frac{1}{2}y_4(\tau)^2-\lambda y_4(\tau)\right)e^{-\tau}d\tau.
\end{align}

We now substitute equations (\ref{eq_inc_b4_obj1})-(\ref{eq_inc_b4_obj2}) in the objective function in (\ref{eq_inc_b4_obj}) to have it written explicitly in terms of the optimization variables, which makes it ready for taking derivatives. Observe, however, that different from the random battery recharges model studied in Section~\ref{sec_rbr}, the Lagrangian in this incremental battery recharges model involves multiple nested integrals, which renders taking derivatives a much more involved operation. For that reason, we refer the reader to Appendix~\ref{apndx_func_der}, in which we summarize some useful results on derivatives under nested integrals that we constantly use in the derivations below.  As noted before, we take derivatives in the following specific alternating order of variables: $y_4(t),~x_3,~y_3(t),~x_2,~y_2(t),~x_1,~y_1(t)$.

Hence, we start now by taking derivative of the Lagrangian with respect to $y_4(t)$ and equate to $0$ to get
\begin{align}
y_4(t)=\lambda+\frac{\gamma_4(t)}{e^{-t}\beta_4(t)},
\end{align}
where the positive term $\beta_4(t)$ is given by 
\begin{align}
&\beta_4(t)\triangleq m_3(\infty,y_1,y_2,t) \nonumber \\
&+\sum_{j=1}^3p_{0,j}\left(e^{x_1}m_2(x_1,y_2,t)\!+\!(p_{1,2}+p_{1,3})e^{x_2}e^{x_1}m_1(x_2,t)\!+\!\left((p_{1,2}\!+\!p_{1,3})p_{2,3}e^{x_2}e^{x_1}\!+\!p_{1,3}e^{x_1}\right)e^{x_3}\right) \nonumber \\
&+\sum_{j=2}^3p_{0,j}\left(e^{x_2}m_1(x_2,t)+p_{2,3}e^{x_2}e^{x_3}\right) \nonumber \\
&+p_{0,3}e^{x_3},
\end{align}
with, according to the notation derived in  Appendix~\ref{apndx_func_der},
\begin{align}
m_3(\infty,y_1,y_2,t)=\int\limits_{\substack{[\infty,~y_1(\tau_1)-\tau_1,~y_2(\tau_1+\tau_2)-(\tau_1+\tau_2)] \\ \tau_1+\tau_2+\tau_3\leq t}} d{\bm \tau}_1^3.
\end{align}
Therefore, $y_4$ is a {\it $\lambda$-threshold policy} given by
\begin{align} \label{eq_inc_b4_y4}
y_4(t)=\begin{cases}\lambda,\quad &t<\lambda \\ 
t,\quad &t\geq\lambda\end{cases}.
\end{align}

Next, we take derivative of the Lagrangian with respect to $x_3$ and equate to $0$ to get
\begin{align} 
\frac{\partial \bar{R}_3-\lambda\bar{L}_3}{\partial x_3}=\frac{\eta_3}{\alpha_3},
\end{align}
where the positive constant $\alpha_3$ is given by
\begin{align}
\alpha_3\triangleq &\sum_{j=1}^3p_{0,j}\left((p_{1,2}+p_{1,3})p_{2,3}e^{x_2}e^{x_1}+p_{1,3}e^{x_1}\right)+\sum_{j=2}^3p_{0,j}p_{2,3}e^{x_2}+p_{0,3},
\end{align}
and
\begin{align}
\frac{\partial \bar{R}_3-\lambda\bar{L}_3}{\partial x_3}=x_3-\lambda+e^{x_3}\int_{\tau=0}^{x_3}\left(\frac{1}{2}y_4(\tau)^2-\lambda y_4(\tau)\right)e^{-\tau}d\tau+\frac{1}{2}x_3^2-\lambda x_3.
\end{align}
Thus, for $x_3>0$ we have $\eta_3=0$ by complementary slackness \cite{boyd} and therefore
\begin{align} \label{eq_inc_b4_x3_kkt}
e^{x_3}\int_{\tau=0}^{x_3}\left(\frac{1}{2}y_4(\tau)^2-\lambda y_4(\tau)\right)e^{-\tau}d\tau+\frac{1}{2}x_3^2-\lambda x_3=\lambda-x_3.
\end{align}
Using (\ref{eq_inc_b4_y4}), the above simplifies to
\begin{align} \label{eq_inc_b4_x3}
x_3=\log\left(\frac{1}{e^{-\lambda}-\frac{1}{2}\lambda^2}\right).
\end{align}
Note that the above equation implies that $x_3>\lambda$.

We now state the following assumption for the upcoming analysis; we verify the assumption in a step-by-step manner as we move further into the characterization of the optimal policy below.
\begin{assumption} \label{asmp_monotone}
The optimal policy for $B=4$ satisfies the following:
\begin{align}
y_3(t)>&\lambda,\quad\forall t, \\
x_2>&x_3, \\
y_2(t)>&x_3,\quad\forall t, \\
x_1>&x_2, \\
y_1(t)>&x_2,\quad\forall t.
\end{align}
\end{assumption}

Continuing with the specified order of taking derivatives, we now take derivative of the Lagrangian with respect to $y_3(t)$ and equate to $0$, and use (\ref{eq_inc_b4_x3_kkt}), to get
\begin{align}
&\left(m_2(\infty,y_1,t)+\sum_{j=1}^3p_{0,j}\left(e^{x_1}m_1(x_1,t)+(p_{1,2}+p_{1,3})e^{x_2}e^{x_1}\right)+\sum_{j=2}^3p_{0,j}e^{x_2}\right) \nonumber \\
&\hspace{.25in}\times \left(\left(y_3(t)-\lambda\right)e^{-y_3(t)}-\left(\frac{1}{2}y_3(t)^2-\lambda y_3(t)\right)e^{-y_3(t)} \right. \nonumber \\
&\left. \hspace{.5in} +\left(\frac{1}{2}y_4\left(y_3(t)^-\right)^2-\lambda y_4\left(y_3(t)^-\right)\right)e^{-y_4\left(y_3(t)^-\right)}+\left(\lambda-x_3\right)e^{-y_3(t)}\right)=\gamma_3(t),
\end{align}
where, according to the notation derived in  Appendix~\ref{apndx_func_der},
\begin{align}
m_2(\infty,y_1,t)=\int\limits_{\substack{[\infty,~y_1(\tau_1)-\tau_1] \\ \tau_1+\tau_2\leq t}} d{\bm \tau}_1^2.
\end{align}
We now use the first premise in Assumption~\ref{asmp_monotone}, namely, $y_3(t)>\lambda,~\forall t$, to conclude by (\ref{eq_inc_b4_y4}) that $y_4\left(y_3(t)^-\right)=y_3(t)$, and hence the above equation simplifies to
\begin{align}
y_3(t)=x_3+\frac{\gamma_3(t)}{e^{-y_3(t)}\beta_3(t)},
\end{align}
where the positive term $\beta_3(t)$ is given by
\begin{align}
\beta_3(t)\triangleq m_2(\infty,y_1,t)+\sum_{j=1}^3p_{0,j}\left(e^{x_1}m_1(x_1,t)+(p_{1,2}+p_{1,3})e^{x_2}e^{x_1}\right)+\sum_{j=2}^3p_{0,j}e^{x_2}.
\end{align}
Therefore, $y_3$ is an {\it $x_3$-threshold policy} given by
\begin{align} \label{eq_inc_b4_y3}
y_3(t)=\begin{cases}x_3,\quad &t<x_3 \\ 
t,\quad &t\geq x_3\end{cases},
\end{align}
which verifies that $y_3(t)>\lambda,~\forall t$, the first premise of Assumption~\ref{asmp_monotone}, since $x_3>\lambda$ from (\ref{eq_inc_b4_x3}).

We now take derivative of the Lagrangian with respect to $x_2$ and equate to $0$, and use (\ref{eq_inc_b4_x3_kkt}), to get
\begin{align}
\frac{\partial \bar{R}_2-\lambda\bar{L}_2}{\partial x_2}=\frac{\eta_2}{\alpha_2},
\end{align}
where the positive term $\alpha_2$ is given by
\begin{align}
\alpha_2\triangleq \sum_{j=1}^3p_{0,j}(p_{1,2}+p_{1,3})e^{x_1}+\sum_{j=2}^3p_{0,j},
\end{align}
and
\begin{align}
&\frac{\partial \bar{R}_2-\lambda\bar{L}_2}{\partial x_2}=x_2-\lambda \nonumber \\
&+ e^{x_2}\left(\int_{\tau=0}^{x_2}\left(\frac{1}{2}y_3(\tau)^2-\lambda y_3(\tau)\right)e^{-y_3(\tau)}d\tau + \hspace{-.3in} \int\limits_{[x_2,y_3(\tau_1)-\tau_1]} \hspace{-.25in} \left(\frac{1}{2}y_4(\tau_1+\tau_2)^2-\lambda y_4(\tau_1+\tau_2)\right)e^{-(\tau_1+\tau_2)}d{\bm \tau}_1^2\right. \nonumber \\
&\left.+\left(\frac{1}{2}y_3(x_2^-)^2-\lambda y_3(x_2^-)\right)e^{-y_3(x_2^-)}+\int_{\tau_2=0}^{y_3(x_2^-)-x_2}\left(\frac{1}{2}y_4(x_2+\tau_2)^2-\lambda y_4(x_2+\tau_2)\right)e^{-(x_2+\tau_2)}d\tau_2\right. \nonumber \\
&\left.+\left(\int_{\tau_2=0}^{y_3(x_2^-)-x_2}e^{-(x_2+\tau_2)}d\tau_2 + p_{2,3}\right)\left(\lambda-x_3\right) \right).
\end{align}
We now use the second premise of Assumption~\ref{asmp_monotone}, namely, $x_2>x_3$, to conclude by (\ref{eq_inc_b4_y3}) that $y_3(x_2^-)=x_2$, and hence the above equation simplifies to
\begin{align}
&\frac{\partial \bar{R}_2-\lambda\bar{L}_2}{\partial x_2}=x_2-\lambda + e^{x_2}\Bigg(\int_{\tau=0}^{x_2}\left(\frac{1}{2}y_3(\tau)^2-\lambda y_3(\tau)\right)e^{-y_3(\tau)}d\tau \nonumber \\
&+\int\limits_{[x_2,y_3(\tau_1)-\tau_1]}\left(\frac{1}{2}y_4(\tau_1+\tau_2)^2-\lambda y_4(\tau_1+\tau_2)\right)e^{-(\tau_1+\tau_2)}d{\bm \tau}_1^2 + p_{2,3} \left(\lambda-x_3\right)\Bigg) + \frac{1}{2}x_2^2-\lambda x_2.
\end{align}
Thus, for $x_2>0$ we have $\eta_2=0$ by complementary slackness \cite{boyd} and therefore
\begin{align} \label{eq_inc_b4_x2_kkt}
&e^{x_2}\Bigg(\int_{\tau=0}^{x_2}\left(\frac{1}{2}y_3(\tau)^2-\lambda y_3(\tau)\right)e^{-y_3(\tau)}d\tau \nonumber \\
&\hspace{.25in}+ \int\limits_{[x_2,y_3(\tau_1)-\tau_1]}\left(\frac{1}{2}y_4(\tau_1+\tau_2)^2-\lambda y_4(\tau_1+\tau_2)\right)e^{-(\tau_1+\tau_2)}d{\bm \tau}_1^2 + p_{2,3} \left(\lambda-x_3\right)\Bigg) + \frac{1}{2}x_2^2-\lambda x_2 \nonumber \\
&=\lambda-x_2.
\end{align}
Using (\ref{eq_inc_b4_y4}) and (\ref{eq_inc_b4_y3}), the above simplifies to
\begin{align} \label{eq_inc_b4_x2}
x_2=&\log\left(\frac{1}{(\lambda+1)e^{-\lambda}+\frac{1}{2}\lambda^2+\lambda-x_3\left(e^{-x_3}+1\right)}\right) \nonumber \\
=&\log\left(\frac{1}{(\lambda+1)e^{-\lambda}+\frac{1}{2}\lambda^2+\lambda+\log\left(e^{-\lambda}-\frac{1}{2}\lambda^2\right)\left(e^{-\lambda}-\frac{1}{2}\lambda^2+1\right)}\right),
\end{align}
where the second equality follows from (\ref{eq_inc_b4_x3}). We note that the above equation has a real-valued solution only if $\lambda\leq0.72$. However, we know from \cite{arafa-age-sgl} that the optimal solution for the $B=2$ case is $0.72$, and hence the optimal solution, i.e., $\lambda^*$, for $B=4$ cannot be larger than $0.72$. We also note that the optimal $\lambda^*$ cannot be smaller than $0.5$, the solution for the $B=\infty$ case reported in \cite{jing-age-online}. Moreover, for $\lambda\in[0.5,0.72]$, it holds that $x_2>x_3$, verifying the second premise in Assumption~\ref{asmp_monotone}.

We now take derivative of the Lagrangian with respect to $y_2(t)$ and equate to $0$ to get
\begin{align} \label{eq_inc_b4_y2_kkt}
&m_1(\infty,t)\Bigg( \left(y_2(t)-\lambda\right)e^{-y_2(t)} - \left(\frac{1}{2}y_2(t)^2-\lambda y_2(t)\right) e^{-y_2(t)} \nonumber \\
&\hspace{.5in}+ \left(\frac{1}{2}y_3\left(y_2(t)^-\right)^2-\lambda y_3\left(y_2(t)^-\right)\right)e^{-y_3\left(y_2(t)^-\right)} \nonumber \\
&\hspace{.5in}+ \int_{\tau_4=0}^{y_3\left(y_2(t)^-\right)-y_2(t)}\frac{1}{2}\left(y_4\left(y_2(t)^-+\tau_4\right)^2-\lambda y_4\left(y_2(t)^-+\tau_4\right)\right)e^{-(y_2(t)+\tau_4)} d\tau_4 \Bigg) \nonumber \\
&+\frac{\partial\sum_{j=1}^3p_{0,j}}{\partial y_2(t)}\left(\mathbb{E}[R_1]-\lambda\mathbb{E}[L_1]\right)+\sum_{j=1}^3p_{0,j}\frac{\partial \mathbb{E}[R_1]-\lambda\mathbb{E}[L_1]}{\partial y_2(t)} \nonumber \\
&+\frac{\partial\sum_{j=2}^3p_{0,j}}{\partial y_2(t)}\left(\bar{R}_2-\lambda\bar{L}_2\right) +\frac{\partial p_{0,3}}{\partial y_2(t)}\left(\bar{R}_3-\lambda\bar{L}_3\right)=\gamma_2(t),
\end{align}
where, according to the notation derived in Appendix~\ref{apndx_func_der},
\begin{align}
m_1(\infty,t)=\int_{\substack{\tau_1=0 \\ \tau_1\leq t}}^\infty d\tau_1=t,
\end{align}
and, using (\ref{eq_inc_b4_x3_kkt}) and (\ref{eq_inc_b4_x2_kkt}),
\begin{align}
\frac{\partial\sum_{j=1}^3p_{0,j}}{\partial y_2(t)}=&m_1(\infty,t)\left(-e^{-y_2(t)}+e^{-y_2(t)}+\int_{\tau_4=0}^{y_3\left(y_2(t)^-\right)-y_2(t)}e^{-(y_2(t)+\tau_4)}d\tau_4\right), \\
\frac{\partial \mathbb{E}[R_1]-\lambda\mathbb{E}[L_1]}{\partial y_2(t)}=&e^{x_1}\Bigg( \left(y_2(t)-\lambda\right)e^{-y_2(t)} - \left(\frac{1}{2}y_2(t)^2-\lambda y_2(t)\right) e^{-y_2(t)} \nonumber \\
&\hspace{.05in}+ \left(\frac{1}{2}y_3\left(y_2(t)^-\right)^2-\lambda y_3\left(y_2(t)^-\right)\right)e^{-y_3\left(y_2(t)^-\right)} \nonumber \\
&\hspace{.05in}+ \int_{\tau_4=0}^{y_3\left(y_2(t)^-\right)-y_2(t)}\!\frac{1}{2}\left(y_4\left(y_2(t)^-+\tau_4\right)^2\!-\!\lambda y_4\left(y_2(t)^-+\tau_4\right)\right)e^{-(y_2(t)+\tau_4)} d\tau_4 \nonumber \\
&\hspace{.05in}+ \left(\lambda-x_2\right)e^{-y_2(t)}\Bigg), \\
\frac{\partial\sum_{j=2}^3p_{0,j}}{\partial y_2(t)}=&m_1(\infty,t)\left(e^{-y_2(t)}+\int_{\tau_4=0}^{y_3\left(y_2(t)^-\right)-y_2(t)}e^{-(y_2(t)+\tau_4)}d\tau_4\right), \\
\bar{R}_2-\lambda\bar{L}_2=&\lambda-x_2, \\
\frac{\partial p_{0,3}}{\partial y_2(t)}=&m_1(\infty,t)\int_{\tau_4=0}^{y_3\left(y_2(t)^-\right)-y_2(t)}e^{-(y_2(t)+\tau_4)}d\tau_4.
\end{align}
We now use the third premise in Assumption~\ref{asmp_monotone}, namely, $y_2(t)>x_3,~\forall t$, to conclude by (\ref{eq_inc_b4_y3}) that $y_3\left(y_2(t)^-\right)=y_2(t)$, and hence the above equations simplify upon substituting in (\ref{eq_inc_b4_y2_kkt}) to
\begin{align}
y_2(t)=x_2+\frac{\gamma_2(t)}{e^{-y_2(t)}\beta_2(t)},
\end{align}
where the positive term $\beta_2(t)$ is given by
\begin{align}
\beta_2(t)\triangleq m_1(\infty,t)+\sum_{j=1}^3p_{0,j}e^{x_1}.
\end{align}
Therefore, $y_2$ is an {\it $x_2$-threshold policy} given by
\begin{align} \label{eq_inc_b4_y2}
y_2(t)=\begin{cases}x_2,\quad &t<x_2 \\ 
t,\quad &t\geq x_2\end{cases},
\end{align}
which verifies that $y_2(t)>x_3,~\forall t$, the third premise of Assumption~\ref{asmp_monotone}, since $x_2>x_3$ from (\ref{eq_inc_b4_x2}) for $\lambda\in[0.5,0.72]$.

We now take derivative of the Lagrangian with respect to $x_1$ and equate to $0$, and use (\ref{eq_inc_b4_x3_kkt}) and (\ref{eq_inc_b4_x2_kkt}), to get
\begin{align}
\frac{\partial \mathbb{E}[R_1]-\lambda \mathbb{E}[L_1]}{\partial x_1}=\frac{\eta_1}{\alpha_1},
\end{align}
where the positive constant $\alpha_1$ is given by
\begin{align}
\alpha_1\triangleq \sum_{j=1}^3p_{0,j},
\end{align}
and
\begin{align}
&\frac{\partial \mathbb{E}[R_1]-\lambda \mathbb{E}[L_1]}{\partial x_1}=x_1-\lambda \nonumber \\
&+e^{x_1}\Bigg(\int_{\tau=0}^{x_1}\left(\frac{1}{2}y_2(\tau)^2-\lambda y_2(\tau)\right)e^{-y_2(\tau)}d\tau \nonumber \\
&\hspace{.25in}+\int\limits_{[x_1,y_2(\tau_1)-\tau_1]}\left(\frac{1}{2}y_3(\tau_1+\tau_2)^2-\lambda y_3(\tau_1+\tau_2)\right)e^{-y_3(\tau_1+\tau_2)}d{\bm \tau}_1^2 \nonumber \\
&\hspace{.25in}+\int\limits_{[x_1,y_2(\tau_1)-\tau_1,y_3(\tau_1+\tau_2)-(\tau_1+\tau_2)]} \hspace{-.5in} \left(\frac{1}{2}y_4(\tau_1+\tau_2+\tau_3)^2-\lambda y_4(\tau_1+\tau_2+\tau_3)\right)e^{-(\tau_1+\tau_2+\tau_3)}d{\bm \tau}_1^3 \Bigg) \nonumber \\
&+e^{x_1}\Bigg(\left(\frac{1}{2}y_2\left(x_1^-\right)^2-\lambda y_2\left(x_1^-\right)\right)e^{-y_2\left(x_1^-\right)} \nonumber \\ 
&\hspace{.25in}+\int_{\tau_2=0}^{y_2\left(x_1^-\right)-x_1}\left(\frac{1}{2}y_3\left(x_1^-+\tau_2\right)^2-\lambda y_3\left(x_1^-+\tau_2\right)\right)e^{-y_3\left(x_1^-+\tau_2\right)}d\tau_2 \nonumber \\
&\hspace{.25in}+\int\limits_{\left[y_2\left(x_1^-\right)-x_1,y_3\left(x_1^-+\tau_2\right)-(x_1+\tau_2)\right]} \hspace{-.5in} \left(\frac{1}{2}y_4\left(x_1^-+\tau_2+\tau_3\right)^2-\lambda y_4\left(x_1^-+\tau_2+\tau_3\right)\right)e^{-(x_1+\tau_2+\tau_3)}d{\bm \tau}_2^3 \Bigg) \nonumber \\
&+(p_{1,2}+p_{1,3})e^{x_1}(\lambda-x_2)+p_{1,3}e^{x_1}(\lambda-x_3) \nonumber \\
&+\left(\int_{\tau_2=0}^{y_2\left(x_1^-\right)-x_1}e^{-y_3\left(x_1^-+\tau_2\right)}d\tau_2+\int\limits_{\left[y_2\left(x_1^-\right)-x_1,y_3\left(x_1^-+\tau_2\right)-(x_1+\tau_2)\right]} \hspace{-.5in}e^{-(x_1+\tau_2+\tau_3)}d{\bm \tau}_2^3\right)e^{x_1}(\lambda-x_2) \nonumber \\
&+\int\limits_{\left[y_2\left(x_1^-\right)-x_1,y_3\left(x_1^-+\tau_2\right)-(x_1+\tau_2)\right]} \hspace{-.5in}e^{-(x_1+\tau_2+\tau_3)}d{\bm \tau}_2^3e^{x_1}(\lambda-x_3).
\end{align}
We now use the fourth premise of Assumption~\ref{asmp_monotone}, namely, $x_1>x_2$, to conclude by (\ref{eq_inc_b4_y2}) that $y_2(x_1^-)=x_1$ and by (\ref{eq_inc_b4_y3}) that $y_3(x_1^-+\tau_2)=x_1+\tau_2,~\forall\tau_2$, and hence the above equation simplifies to
\begin{align}
\frac{\partial \mathbb{E}[R_1]-\lambda \mathbb{E}[L_1]}{\partial x_1}=&x_1-\lambda \nonumber \\
&+e^{x_1}\left(\int_{\tau=0}^{x_1}\left(\frac{1}{2}y_2(\tau)^2-\lambda y_2(\tau)\right)e^{-y_2(\tau)}d\tau\right. \nonumber \\ 
&\hspace{.15in}\left.+\int\limits_{[x_1,y_2(\tau_1)-\tau_1]} \hspace{-.3in} \left(\frac{1}{2}y_3(\tau_1+\tau_2)^2-\lambda y_3(\tau_1+\tau_2)\right)e^{-y_3(\tau_1+\tau_2)}d{\bm \tau}_1^2\right. \nonumber \\
&\hspace{.15in}\left.+\int\limits_{[x_1,y_2(\tau_1)-\tau_1,y_3(\tau_1+\tau_2)-(\tau_1+\tau_2)]} \hspace{-.85in} \left(\!\frac{1}{2}y_4(\tau_1+\tau_2+\tau_3)^2\!-\!\lambda y_4(\tau_1+\tau_2+\tau_3)\!\right)\!e^{-(\tau_1+\tau_2+\tau_3)}d{\bm \tau}_1^3 \right. \nonumber \\
&\left. \hspace{.15in} +(p_{1,2}+p_{1,3})(\lambda-x_2)+p_{1,3}(\lambda-x_3) \right) \nonumber \\
&+\frac{1}{2}x_1^2-\lambda x_1.
\end{align}
Thus, for $x_2>0$ we have $\eta_2=0$ by complementary slackness \cite{boyd} and therefore
\begin{align} \label{eq_inc_b4_x1_kkt}
&e^{x_1}\left(\int_{\tau=0}^{x_1}\left(\frac{1}{2}y_2(\tau)^2-\lambda y_2(\tau)\right)e^{-y_2(\tau)}d\tau\right. \nonumber \\ 
&\hspace{.25in}\left.+\int\limits_{[x_1,y_2(\tau_1)-\tau_1]}\left(\frac{1}{2}y_3(\tau_1+\tau_2)^2-\lambda y_3(\tau_1+\tau_2)\right)e^{-y_3(\tau_1+\tau_2)}d{\bm \tau}_1^2\right. \nonumber \\
&\hspace{.25in}\left.+\int\limits_{[x_1,y_2(\tau_1)-\tau_1,y_3(\tau_1+\tau_2)-(\tau_1+\tau_2)]} \hspace{-.5in} \left(\frac{1}{2}y_4(\tau_1+\tau_2+\tau_3)^2-\lambda y_4(\tau_1+\tau_2+\tau_3)\right)e^{-(\tau_1+\tau_2+\tau_3)}d{\bm \tau}_1^3 \right. \nonumber \\
&\left. \hspace{.25in} +(p_{1,2}+p_{1,3})(\lambda-x_2)+p_{1,3}(\lambda-x_3) \right)+\frac{1}{2}x_1^2-\lambda x_1=\lambda-x_1.
\end{align}
Using (\ref{eq_inc_b4_y4}), (\ref{eq_inc_b4_y3}), and (\ref{eq_inc_b4_y2}), the above simplifies to
\begin{align} \label{eq_inc_b4_x1}
x_1=&\log\left(\frac{1}{\left(\frac{1}{2}\lambda^2+3\lambda+6\right)e^{-\lambda}+2\lambda-\frac{1}{2}\lambda^2-x_2-\left(x_2+2\right)e^{-x_2}-x_3-\left(\frac{1}{2}x_3^2+2x_3+3\right)e^{-x_3}}\right) \nonumber \\
=&\log\left(\frac{1}{\left(\frac{1}{2}\lambda^2+\lambda+1\right)e^{-\lambda}-x_2\left(e^{-x_2}+1\right)-x_3\left(\frac{1}{2}x_3e^{-x_3}-1\right)}\right),
\end{align}
where the second equality follows from (\ref{eq_inc_b4_x3}) and (\ref{eq_inc_b4_x2}). We note that the above equation admits a real-valued solution only if $\lambda\leq0.64$. Moreover, for $\lambda\in[0.5,0.64]$, it holds that $x_1>x_2$. Thus, to verify the fourth premise of Assumption~\ref{asmp_monotone}, we need to show that the optimal $\lambda^*\leq0.64$ for $B=4$, which we indeed show towards the end of the analysis.

We finally take derivative of the Lagrangian with respect to $y_1(t)$ and equate to $0$, and use (\ref{eq_inc_b4_x3_kkt}), (\ref{eq_inc_b4_x2_kkt}), and (\ref{eq_inc_b4_x1_kkt}), to get 
\begin{align}
&\left(y_1(t)-\lambda\right)e^{-y_1(t)}-\left(\frac{1}{2}y_1(t)^2-\lambda y_1(t)\right)e^{-y_1(t)} +\left(\frac{1}{2}y_2\left(y_1(t)^-\right)^2-\lambda y_2\left(y_1(t)^-\right)\right)e^{-y_2\left(y_1(t)^-\right)} \nonumber \\
&\int_{\tau_3=0}^{y_2\left(y_1(t)^-\right)-y_1(t)}\left(\frac{1}{2}y_3\left(y_1(t)^-+\tau_3\right)^2-\lambda y_3\left(y_1(t)^-+\tau_3\right)\right)e^{-(y_1(t)+\tau_3)} d\tau_3 \nonumber \\
&+\int\limits_{\left[y_2\left(y_1(t)^-\right)-y_1(t),y_3\left(y_1(t)^-+\tau_3\right)-(y_1(t)+\tau_3)\right]} \hspace{-1in} \left(\frac{1}{2}y_4\left(y_1(t)^-+\tau_3+\tau_4\right)^2-\lambda y_4\left(y_1(t)^-+\tau_3+\tau_4\right)\right)e^{-(y_1(t)+\tau_2+\tau_4)}d{\bm \tau}_3^4 \nonumber \\
&+\left(e^{-y_2\left(y_1(t)^-\right)} + \int_{\tau_3=0}^{y_2\left(y_1(t)^-\right)-y_1(t)}e^{-(y_1(t)+\tau_3)} d\tau_3 + \hspace{-1in} \int\limits_{\left[y_2\left(y_1(t)^-\right)-y_1(t),y_3\left(y_1(t)^-+\tau_3\right)-(y_1(t)+\tau_3)\right]} \hspace{-1in} e^{-(y_1(t)+\tau_2+\tau_4)}d{\bm \tau}_3^4\right)(\lambda-x_1)=\gamma_1(t).
\end{align}
We now use the fifth and final premise in Assumption~\ref{asmp_monotone}, namely, $y_1(t)>x_2,~\forall t$, to conclude by (\ref{eq_inc_b4_y2}) that $y_2\left(y_1(t)^-\right)=y_1(t)$ and by (\ref{eq_inc_b4_y3}) that $y_3\left(y_1(t)^-+\tau_3\right)=y_1(t)+\tau_3,~\forall\tau_3$, and hence the above equation simplifies to
\begin{align}
y_1(t)=x_1+\frac{\gamma_1(t)}{e^{-y_1(t)}}.
\end{align}
Therefore, $y_1$ is an {\it $x_1$-threshold policy} given by
\begin{align} \label{eq_inc_b4_y1}
y_1(t)=\begin{cases}x_1,\quad &t<x_1 \\ 
t,\quad &t\geq x_1\end{cases}.
\end{align}
The above verifies that $y_1(t)>x_2,~\forall t$, the fifth premise of Assumption~\ref{asmp_monotone}, only if $x_1>x_2$ is verified, or equivalently if $\lambda^*\leq0.64$ as discussed after equation (\ref{eq_inc_b4_x1}). We show that this is indeed true by evaluating the optimal policy below.

\begin{figure}[t]
\center
\includegraphics[scale=.5]{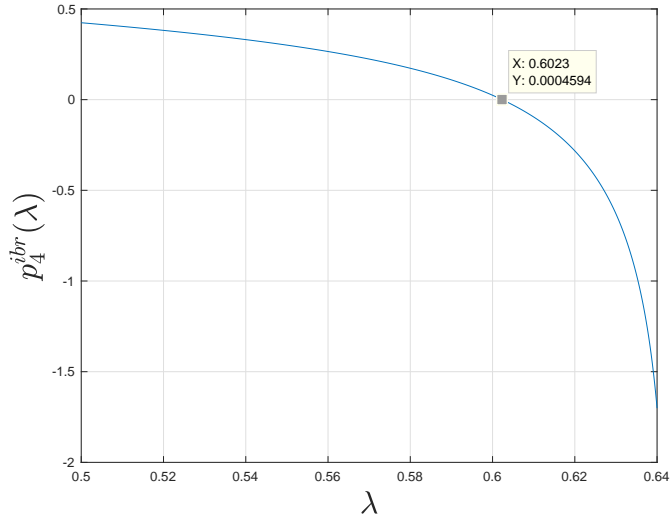}
\caption{$p_4^{\text{ibr}}(\lambda)$ versus $\lambda$.}
\label{fig_p4_vs_lamda}
\end{figure}

We do so by basically substituting the optimal values of the optimization variables, in terms of $\lambda$, in the objective function to evaluate $p_4^{\text{ibr}}(\lambda)$. We then perform a bisection search over $\lambda$ to find the optimal $\lambda^*$ that makes $p_4^{\text{inc}}\left(\lambda^*\right)=0$. As noted in the analysis, we know from \cite{jing-age-online} and \cite{arafa-age-sgl} that $\lambda^*\in[0.5,0.72]$. We have yet to show that $\lambda^*\leq0.64$ to verify the fourth and fifth premises of Assumption~\ref{asmp_monotone}. After some involved simplifications, which we omit for brevity, we get that $p_4^{\text{ibr}}(\lambda)$ is given by
\begin{align}
p_4^{\text{ibr}}(\lambda)=&e^{-\lambda}\left(\frac{1}{6}\lambda^3+\frac{3}{2}\lambda^2+6\lambda+10\right)-\frac{1}{2}\lambda^2 \nonumber \\
&-(x_1-\lambda)-(x_2-\lambda)-(x_3-\lambda) \nonumber \\
&-(x_1+2)e^{-x_1}-\left(\frac{1}{2}x_2^2+2x_2+3\right)e^{-x_2}-\left(\frac{1}{6}x_3^3+x_3^2+3x_3+4\right)e^{-x_3},
\end{align}
with $x_3$, $x_2$, and $x_1$ given by (\ref{eq_inc_b4_x3}), (\ref{eq_inc_b4_x2}), and (\ref{eq_inc_b4_x1}), respectively. In Fig.~\ref{fig_p4_vs_lamda}, we plot $p_4^{\text{ibr}}(\lambda)$ versus $\lambda$. We see that that the optimal $\lambda^*\approx0.6023$, whence the fourth and fifth premises of Assumption~\ref{asmp_monotone} are verified, with $x_3^*\approx1.005$, $x_2^*\approx1.243$, and $x_1^*\approx1.636$.

We note that Assumption~\ref{asmp_monotone} has an intuitive explanation; it basically says that the sensor is less eager to send an update when it has relatively lower energy available in its battery than it is when it has relatively higher energy available.

In summary, given the values of the thresholds $\lambda^*$, $x_3^*$, $x_2^*$ and $x_1^*$ above, the sensor uses each of them to determine whether to send a new status update by comparing the AoI to the threshold corresponding to the amount of energy available: $\lambda^*$ for full battery, and $x_j^*$ for $1\leq j\leq3$ energy units. We finally note that while we work in this section with $B=4$, the methodology adopted to characterize the optimal threshold policy in closed-form works for general $B>1$. As mentioned earlier, working with $B=4$ strikes a balance between simplicity of presentation and revealing the minute details of the analysis.

%============
\section{Numerical Evaluations}

\begin{figure}[t]
\center
\includegraphics[scale=.5]{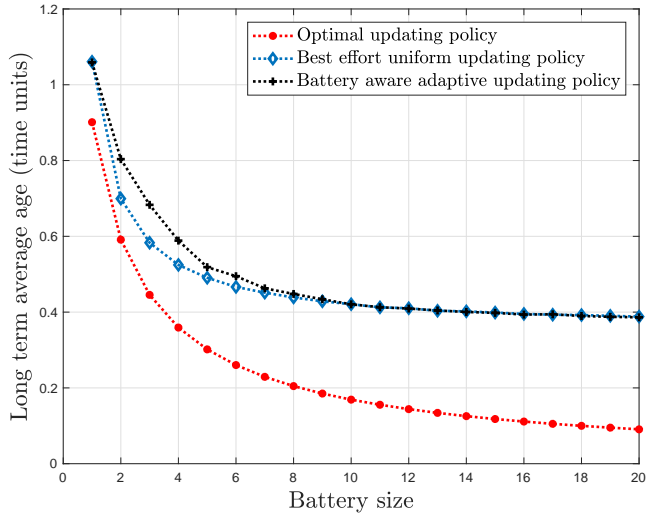}
\caption{Comparison of long term average age versus battery size under different update policies for the RBR model.}
\label{fig_age_battery_rbr}
\end{figure}

In this section, we present some numerical examples for both the RBR and the IBR models. We compare the optimal policy with two other update policies. The first is a best effort uniform updating policy, where the sensor aims at sending an update every $1/\nu$, with $\nu$ representing the average recharging rate, only if it has energy available, and stays silent otherwise. We note that $\nu$ is equal to $B$ in the RBR model, and is equal to $1$ in the IBR model. The rationale is that $1/\nu$ represents the average inter-arrival time between unit arrivals, by which the sensor aims at uniformly spreading its updates over time. The other policy is a slight variation of the battery-aware adaptive update policy proposed in \cite{jing-age-online}, in which the sensor aims at sending its next update depending on the status of its battery: if the battery has more (resp. less) than $B/2$ units, the sensor aims at sending the next update after $1/\nu(1+\beta)$ (resp. $1/\nu(1-\beta)$) time units; and if the battery has exactly $B/2$ units, then the sensor aims at sending the next update after $1/\nu$ time units. We choose $\beta=\log(B)/B$ \cite{jing-age-online}. 

In Fig.~\ref{fig_age_battery_rbr}, we plot the long term average age of the optimal policy in addition to the above two policies, for the RBR model. We consider a system with $T=1000$ time units, and compute the long term average age over $1000$ iterations. We see from the figure that the optimal updating policy outperforms both the uniform and the battery aware adaptive updating policies, and that the gap between them grows larger with the battery size.

\begin{figure}[t]
\center
\includegraphics[scale=.5]{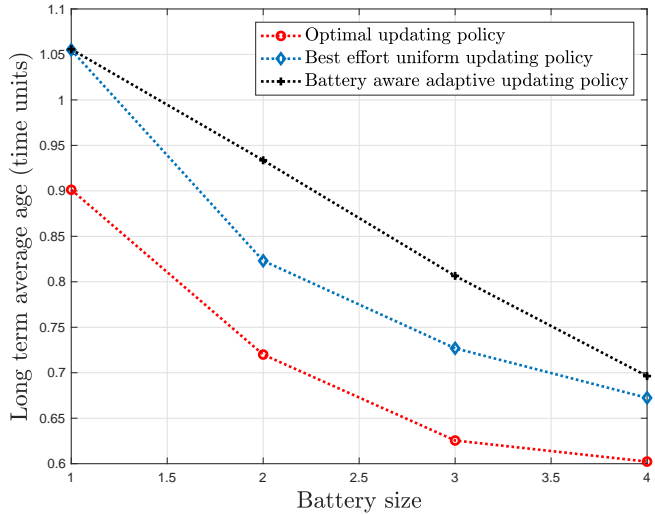}
\caption{Comparison of long term average age versus battery size under different update policies for the IBR model.}
\label{fig_age_battery_ibr}
\end{figure}

We repeat the above for the IBR model, and plot the results in Fig.~\ref{fig_age_battery_ibr}. Again, we observe the superiority of the optimal policy on the other two policies. In this case, however, the gap between the policies shrinks, since all policies converge to $0.5$, the optimal policy for the infinite battery case \cite{jing-age-online}, as the battery size grows large. 

We conclude our numerical results by evaluating the optimal threshold policies derived in this work under an energy arrival model that is different from Poisson. Specifically, we consider a first-order discrete time Markov energy arrival process, which can be at two states: OFF and ON, during a time slot. When the process is in the ON (resp.~OFF) state, one energy unit (resp.~no energy) arrives at the sensor's battery. The process switches from ON to OFF with probability $q_0$, and from OFF to ON with probability $q_1$. This directly leads to having the steady state probability of being in the ON state to be $\frac{q_0}{q_0+q_1}$, and the expected energy arrival value at steady state also given by $\frac{q_0}{q_0+q_1}$ energy units. In order to compare with the unit rate Poisson process that we consider in this work, we choose the time slot duration of the Markovian process to be $\frac{q_0}{q_0+q_1}$ time units, which makes the average recharge rate $B$ energy units (resp.~$1$ energy unit) per unit time for the RBR (resp.~IBR) model, regardless of the values of $q_0$ and $q_1$. Observe that for relatively small values of $q_0$ and $q_1$, the energy arrival process becomes bursty; once it switches to OFF, it stays for a relatively long period of time, after which it switches to ON, and charges the sensor's battery also for a relatively long period of time. While for relatively large values of $q_0$ and $q_1$, the charging process becomes more uniform over time, switching from OFF to ON and vice versa relatively often. We note that such comparison has been carried out for the $B=\infty$ and $B=1$ cases in \cite{jing-age-online}.

In Table~\ref{tab_mrkv}, we list the long term average age achieved by the threshold policies derived in this paper under the Markovian energy arrival process described above. We set $q_0=q_1\triangleq q$ and vary $q$. With the exception of the bursty arrivals case when $q=0.1$, we see that for the other cases the achieved age under Markovian arrivals is relatively low. In particular, for the IBR case the results are very close to the $0.5$ lower bound \cite{jing-age-online}, for $q=0.5$ and $q=1$. Similar conclusions follow for the $B=1$ case. This indicates that while Poisson arrivals allowed optimal theoretical derivations of status update threshold policies, such policies may perform relatively well under general energy arrival models.

\begin{table}
\center
\begin{tabular}[t]{l*{6}{c}r}
Setting		& $B=1$ & $B=4$ (RBR) & $B=4$ (IBR)\\
\hline
Poisson		& 0.9012 & 0.3592 & 0.6023\\
Markov $q=0.1$		& 2.446 & 1.664 & 1.517\\
Markov $q=0.5$		& 0.6916 & 0.2804 & 0.5354\\
Markov $q=1$		& 0.4992 & 0.2517 & 0.5002
\end{tabular}
\caption{Long term average age achieved by threshold policies under Markovian energy arrivals versus Poisson.}
\label{tab_mrkv}
\end{table}

%============
\section{Conclusion and Discussion}

The optimality of online threshold status update policies has been shown for an energy harvesting sensor with a finite battery, and zero service times. We have considered two energy recharging models: RBR, and IBR. In both models, energy arrives according to a Poisson process with unit rate, at times that are only revealed causally over time, yet with amounts that fully recharge the battery in the RBR model, and with unit amounts in the IBR model. For both models, we have shown that the optimal status update policy has a renewal structure, in which update times follow a specific renewal process depending on the recharging model, and then have shown that the optimal renewal policy is an energy-dependent threshold policy, where an update is sent only if the AoI grows above a certain threshold that depends on the energy available. The optimal thresholds have been explicitly characterized in terms of the optimal age, which has in turn been found via a bisection search over a bounded interval that is strictly contained inside the unit interval. The results have shown that, for both recharging models, the optimal thresholds are monotonically decreasing as a function of the energy available in the battery, and that the smallest threshold, when the battery is full, is equal to the optimal long term average AoI.

We note that although the paper addresses an online energy arrival setting, in which the sensor needs to decide on when to send a new update on the fly, all the computations can be carried offline. That is, computing the optimal values of the thresholds for both the RBR and IBR models can be done before the communication session starts, based only on the average arrival rate and battery size. Such threshold policies are not only optimal, they are also relatively simple to implement; the sensor only needs to compare the elapsed time to some threshold before deciding on sending a new update.

We conclude by discussing some possible extensions of the ideas in this paper. One extension is to study the problem in which updates are subject to erasures, and show how threshold policies behave under such setting. We note that an effort toward that has been made for the setting in which the sensor is equipped with a unit-sized battery in \cite{arafa-age-erasure-no-fb, arafa-age-erasure-fb}, where erasure-dependent threshold policies are shown to be age-minimal. Another extension would be to combine both recharging models studied in this paper in one setting where energy also arrives according to a Poisson process, yet with value $e\in\{1,2,\dots,B\}$ with some probability mass function on the set $\{1,2,\dots,B\}$. Generally, it would be of interest to study the effects of energy arrival processes other than Poisson, that do not possess the memorylessness property of the inter-arrival times, on the optimal policy, and whether threshold policies are optimal under more general settings. Another interesting setting is the case in which some updates may have higher priorities, in the sense of having higher age penalty, and therefore allocating more energy resources toward them may be more beneficial. Finally, although the threshold policy in itself is relatively simple, the proof of its optimality and its analysis, especially in the IBR model, are rather involved. Hence, it would be of interest to analyze the performance of threshold policies and other forms of policies, especially if erasures are included or if general energy arrival models are considered, and show their near-optimality with respect to the optimal solution, in the same sense of the online energy harvesting literature in \cite{ozgur_online_su, ozgur_online_mac, baknina_online_mac, baknina_online_bc, baknina-online-proc, varan-online-ffp-cognitive, baknina-online-data, ozgur-online-gen, arafa-baknina-ffp, arafa-mobile-tgcn, ozgur-online-block-iid}.

%================================
\section{Appendix}

\subsection{Proof of Theorem~\ref{thm_rbr_rnwl}} \label{apndx_rbr_rnwl}

Consider any feasible uniformly bounded policy. Let ${\bm x}_i\triangleq\{x_{1,i},\dots,x_{k,i}\}$, and let us denote by $R\left({\bm x}_i\right)$ the area under the age curve during the $i$th epoch. Then
\begin{align}
R\left({\bm x}_i\right)=&\frac{1}{2}\sum_{j=1}^{k-1}\left(x_{j,i}-x_{j+1,i}\right)^2\mathbbm{1}_{x_{j,i}\leq\tau_i}+\frac{1}{2}\left(x_{k,i}\left(\tau_i\right)-\sum_{j=1}^{k-1}x_{j,i}\mathbbm{1}_{x_{j,i}\leq\tau_i<x_{j-1,i}}\right)^2,
\end{align} 
where $\mathbbm{1}_A$ equals 1 if the event $A$ is true, and 0 otherwise. Next, for a given time $T$, let $N_T$ denote the number of epochs that have already {\it started} by time $T$, and for a fixed history $\mathcal{H}_{i-1}$, let us group all the status updating sample paths that have the same $\tau_i$ and perform a statistical averaging over all of them to get the following average age in the $i$th epoch:
\begin{align}
\hat{R}_i\left(\gamma,\mathcal{H}_{i-1}\right)\triangleq\mathbb{E}\left[R\left({\bm x}_i\right)|\tau_i=\gamma,\mathcal{H}_{i-1}\right].
\end{align}
Then, we have
\begin{align}
\mathbb{E}&\left[R\left({\bm x}_i\right)\mathbbm{1}_{i\leq N_T}\right] =\mathbb{E}_{\mathcal{H}_{i-1}}\left[\mathbb{E}_{\tau_i}\left[\hat{R}_i\left(\gamma,\mathcal{H}_{i-1}\right)\right]\cdot\mathbbm{1}_{i\leq N_T}\Big|\mathcal{H}_{i-1}\right],\label{eq_rbr_pf_ren_1}
\end{align}
where equality follows since $\mathbbm{1}_{i\leq N_T}$ is independent of $\tau_i$ given $\mathcal{H}_{i-1}$. Similarly, define the average $i$th epoch length as
\begin{align}
\hat{x}_{k,i}\left(\gamma,\mathcal{H}_{i-1}\right)\triangleq\mathbb{E}\left[x_{k,i}|\tau_i=\gamma,\mathcal{H}_{i-1}\right].
\end{align}

Next, note that by (\ref{eq_aoi}), the following holds:
\begin{align} \label{eq_rbr_aoi_bd}
\frac{1}{T}\sum_{i=1}^\infty R_i\mathbbm{1}_{i\leq N_T-1}\leq \frac{r(T)}{T} \leq \frac{1}{T}\sum_{i=1}^\infty R_i\mathbbm{1}_{i\leq N_T}.
\end{align} 
Following similar analysis to that in \cite[Appendix C-1]{jing-age-online}, one can show that $\lim_{T\rightarrow\infty} \frac{\mathbb{E}\left[R_{N_T}\right]}{T}=0$
for any uniformly bounded policy. Hence, the expected values of the upper and lower bounds in (\ref{eq_rbr_aoi_bd}) are equal as $T\rightarrow\infty$. Hence, in the sequel, we derive a lower bound on $\frac{1}{T}\mathbb{E}\left[\sum_{i=1}^\infty R_i\mathbbm{1}_{i\leq N_T}\right]$ and use the above note to conclude that it is also a lower bound on $\frac{\mathbb{E}\left[r(T)\right]}{T}$ as $T\rightarrow\infty$. Towards that end, note that $\mathbb{E}\left[\sum_{i=1}^\infty x_{k,i}\mathbbm{1}_{i\leq N_T}\right]\geq T$. Then, we have
\begin{align}
\frac{1}{T}\mathbb{E}\left[\sum_{i=1}^\infty R\left({\bm x}_i\right)\mathbbm{1}_{i\leq N_T}\right] \geq \frac{\mathbb{E}\left[\sum_{i=1}^\infty R\left({\bm x}_i\right)\mathbbm{1}_{i\leq N_T}\right]}{\mathbb{E}\left[\sum_{i=1}^\infty x_{k,i}\mathbbm{1}_{i\leq N_T}\right]}.
\end{align}
Next, we proceed by lower bounding the right hand side of the above equation through a series of equations as follows:
\begin{align}
\frac{\mathbb{E}\left[\sum_{i=1}^\infty R\left({\bm x}_i\right)\mathbbm{1}_{i\leq N_T}\right]}{\mathbb{E}\left[\sum_{i=1}^\infty x_{k,i}\mathbbm{1}_{i\leq N_T}\right]} &=\frac{\sum_{i=1}^\infty\mathbb{E}_{\mathcal{H}_{i-1}}\left[\mathbb{E}_{\tau_i}\left[\hat{R}_i\left(\gamma,\mathcal{H}_{i-1}\right)\right]\cdot\mathbbm{1}_{i\leq N_T}\Big|\mathcal{H}_{i-1}\right]}{\mathbb{E}\left[\sum_{i=1}^\infty x_{k,i}\mathbbm{1}_{i\leq N_T}\right]} \label{eq_rbr_pf_ren_2} \\
&=\frac{\sum_{i=1}^\infty\mathbb{E}_{\mathcal{H}_{i-1}}\left[\mathbb{E}_{\tau_i}\left[\hat{x}_{k,i}\left(\gamma,\mathcal{H}_{i-1}\right)\right]
\cdot \frac{\mathbb{E}_{\tau_i}\left[\hat{R}_i\left(\gamma,\mathcal{H}_{i-1}\right)\right]}{\mathbb{E}_{\tau_i}\left[\hat{x}_{k,i}\left(\gamma,\mathcal{H}_{i-1}\right)\right]} 
\cdot \mathbbm{1}_{i\leq N_T}\Big|\mathcal{H}_{i-1}\right]}{\mathbb{E}\left[\sum_{i=1}^\infty x_{k,i}\mathbbm{1}_{i\leq N_T}\right]} \\
&\geq\frac{\sum_{i=1}^\infty\mathbb{E}_{\mathcal{H}_{i-1}}\left[\mathbb{E}_{\tau_i}\left[\hat{x}_{k,i}\left(\gamma,\mathcal{H}_{i-1}\right)\right]
\cdot R^*\left(\mathcal{H}_{i-1}\right)
\cdot \mathbbm{1}_{i\leq N_T}\Big|\mathcal{H}_{i-1}\right]}{\mathbb{E}\left[\sum_{i=1}^\infty x_{k,i}\mathbbm{1}_{i\leq N_T}\right]} \\
&\geq R_{\min},
\end{align}
where (\ref{eq_rbr_pf_ren_2}) follows from (\ref{eq_rbr_pf_ren_1}) and the monotone convergence theorem, $R^*\left(\mathcal{H}_{i-1}\right)$ is the minimum value of $\frac{\mathbb{E}_{\tau_i}\left[\hat{R}_i\left(\gamma,\mathcal{H}_{i-1}\right)\right]}{\mathbb{E}_{\tau_i}\left[\hat{x}_{k,i}\left(\gamma,\mathcal{H}_{i-1}\right)\right]}$, and $R_{\min}$ is the minimum value of $R^*\left(\mathcal{H}_{i-1}\right)$ over all possible epochs and their corresponding histories, i.e., the minimum over all $i$ and $\mathcal{H}_{i-1}$.

Observe that a policy achieving $R^*\left(\mathcal{H}_{i-1}\right)$ is a policy where $\{x_{j,i}\}_{j=1}^{k-1}$ are constants and $x_{k,i}$ is a function of $\tau_i$ only, since the history $\mathcal{H}_{i-1}$ is fixed. Now, if we repeat the policy that achieves $R_{\min}$ over all epochs, we get a renewal policy where $\forall i$ $\{x_{j,i}\}_{j=1}^{k-1}$ are constants and $x_{k,i}$ is only a function of $\tau_i$. Since $\tau_i$'s are i.i.d., the epoch lengths are also i.i.d., and $\{l_i\}$ forms a renewal process. This completes the proof.

\subsection{Proof of Theorem~\ref{thm_inc_rnwl}} \label{apndx_inc_rnwl}

We prove this by showing that any given status update policy that is uniformly bounded according to Definition~\ref{def_ubp} is outperformed by a renewal policy as defined in the theorem. The essence of the proof is similar to that of Theorem~\ref{thm_rbr_rnwl}, yet with some important different details. Let us consider the $i$th epoch (time between two consecutive visits to state $(k,0)$); we introduce the following notation regarding the energy arrivals occurring in it. Let $\tau_{1,i}$ denote the time until the first energy arrival after the epoch starts, and let there be $j_1$ status updates after that energy arrival {\it before} a second energy arrival occurs. If $j_1\geq1$, then let $\tau_{2,i}$ denote the time until the first energy arrival {\it after} the $j_1$th update. Otherwise, if $j_1=0$, then let $\tau_{2,i}$ denote the inter-arrival time between the first and the second energy arrivals in the epoch. Similarly, let there be $j_2$ status updates after the second energy arrival {\it before} a third energy arrival occurs. If $j_2\geq1$, then let $\tau_{3,i}$ denote the time until the first energy arrival {\it after} the $j_2$th update. Otherwise, if $j_2=0$, then let $\tau_{3,i}$ denote the inter-arrival time between the second and the third energy arrivals in the epoch. We continue defining $\tau_{j,i}$'s, $j=1,2,\dots$, until the epoch ends by returning back to state $(k,0)$ again. Finally, in the event that the $j$th energy arrival in the epoch makes the battery full, then we wait until the first status update occurs after that event and denote by $\tau_{j+1,i}$ the time until the first energy arrival {\it after} that update, i.e., we do not account for energy arrivals that cause battery overflows.

As noted before Theorem~\ref{thm_inc_rnwl}, there can possibly be an infinite number of updates before the system returns back to state $(k,0)$, depending on the energy arrival pattern and the update time decisions. For a given status update policy, one can enumerate all such patterns. For instance, following the above notation, the first pattern could be when the system goes from state $(k,0)$ to state $(k+1,\tau_{1,i})$ and then to state $(k,0)$ again; the second pattern could be when the system goes through the following sequence of states: $(k,0)-(k+1,\tau_{1,i})-(k+2,\tau_{1,i}+\tau_{2,i})-(k+1,0)-(k,0)$; and so on. Let the vector ${\bm \tau}_{m,i}$ contain all the $\tau_{j,i}$'s in the $m$th pattern. Note that this vector's length varies with the pattern. For instance, we have ${\bm \tau}_{1,i}=\tau_{1,i}$ and ${\bm \tau}_{2,i}=[\tau_{1,i},\tau_{2,i}]$ for the above two pattern examples, respectively. For a given status update policy, one can also compute the probability of occurrence of the $m$th pattern in the $i$th epoch, denoted by $p_{m,i}$, with $\sum_{m=1}^\infty p_{m,i}=1$. Let us also denote by $R_{m,i}$ the area under the age curve in that epoch, given that it went through the $m$th pattern.

Next, for a fixed history $\mathcal{H}_{i-1}$ and a pattern $m$, let us group all the status updating sample paths that have the same ${\bm \tau}_{m,i}$ and perform a statistical averaging over all of them to get the following average age in the $i$th epoch given that it went through the $m$th pattern:
\begin{align}
\hat{R}_{m,i}\left({\bm \gamma}_m,\mathcal{H}_{i-1}\right)\triangleq\mathbb{E}\left[R_{m,i}|{\bm \tau}_{m,i}={\bm \gamma}_m,\mathcal{H}_{i-1}\right].
\end{align}
Now for a given time $T$, let $N_T$ denote the number of epochs that have already {\it started} by time $T$. Then, we have
\begin{align}
\mathbb{E}\left[R_{m,i}\cdot\mathbbm{1}_{i\leq N_T}\right] =\mathbb{E}_{\mathcal{H}_{i-1}}\left[\mathbb{E}_{{\bm \tau}_{m,i}}\left[\hat{R}_{m,i}\left({\bm \gamma}_m,\mathcal{H}_{i-1}\right)\right]\cdot\mathbbm{1}_{i\leq N_T}\Big|\mathcal{H}_{i-1}\right], \label{eq_inc_pf_ren_1}
\end{align}
where equality follows since $\mathbbm{1}_{i\leq N_T}$ is independent of ${\bm \tau}_{m,i}$ given $\mathcal{H}_{i-1}$. Similarly, let $x_{k,m,i}$ denote the length of the $i$th epoch under the $m$th pattern, and define its (conditional) average as
\begin{align}
\hat{x}_{k,m,i}\left({\bm \gamma}_m,\mathcal{H}_{i-1}\right)\triangleq\mathbb{E}\left[x_{k,m,i}|{\bm \tau}_{m,i}={\bm \gamma}_m,\mathcal{H}_{i-1}\right].
\end{align}
Finally, we denote by $R_i$ and $x_{k,i}$ the area under the age curve in the $i$th epoch and its length, respectively, irrespective of which pattern it went through.

Next, note that by (\ref{eq_aoi}), the following holds
\begin{align} \label{eq_inc_aoi_bd}
\frac{1}{T}\sum_{i=1}^\infty R_i\mathbbm{1}_{i\leq N_T-1}\leq \frac{r(T)}{T} \leq \frac{1}{T}\sum_{i=1}^\infty R_i\mathbbm{1}_{i\leq N_T}.
\end{align} 
Following similar analysis as in \cite[Appendix C-1]{jing-age-online}, one can show that $\lim_{T\rightarrow\infty} \frac{\mathbb{E}\left[R_{N_T}\right]}{T}=0$
for any uniformly bounded policy as in Definition~\ref{def_ubp}. Hence, the expected values of the upper and lower bounds in (\ref{eq_inc_aoi_bd}) are equal as $T\rightarrow\infty$. Hence, in the sequel, we derive a lower bound on $\frac{1}{T}\mathbb{E}\left[\sum_{i=1}^\infty R_i\mathbbm{1}_{i\leq N_T}\right]$ and use the above note to conclude that it is also a lower bound on $\frac{\mathbb{E}\left[r(T)\right]}{T}$ as $T\rightarrow\infty$. Towards that end, note that $\mathbb{E}\left[\sum_{i=1}^\infty x_{k,i}\mathbbm{1}_{i\leq N_T}\right]\geq T$. Then, we have
\begin{align}
\frac{1}{T}\mathbb{E}\left[\sum_{i=1}^\infty R_i\mathbbm{1}_{i\leq N_T}\right] \geq \frac{\mathbb{E}\left[\sum_{i=1}^\infty R_i\mathbbm{1}_{i\leq N_T}\right]}{\mathbb{E}\left[\sum_{i=1}^\infty x_{k,i}\mathbbm{1}_{i\leq N_T}\right]}.
\end{align}
We now proceed by lower bounding the right hand side of the above equation through a series of equations as follows:
\begin{align}
&\frac{\mathbb{E}\left[\sum_{i=1}^\infty R_i\mathbbm{1}_{i\leq N_T}\right]}{\mathbb{E}\left[\sum_{i=1}^\infty x_{k,i}\mathbbm{1}_{i\leq N_T}\right]} =\frac{\sum_{i=1}^\infty \mathbb{E}_{\mathcal{H}_{i-1}}\left[ \sum_{m=1}^\infty p_{m,i} \mathbb{E}_{{\bm \tau}_{m,i}}\left[\hat{R}_{m,i}\left({\bm \gamma}_m,\mathcal{H}_{i-1}\right)\right]\cdot\mathbbm{1}_{i\leq N_T}\Big|\mathcal{H}_{i-1}\right]}{\mathbb{E}\left[\sum_{i=1}^\infty x_{k,i}\mathbbm{1}_{i\leq N_T}\right]} \label{eq_inc_pf_ren_2} \\
&=\frac{\sum_{i=1}^\infty \mathbb{E}_{\mathcal{H}_{i-1}}\left[ \sum_{m=1}^\infty p_{m,i} \mathbb{E}_{{\bm \tau}_{m,i}}\left[\hat{x}_{k,m,i}\left({\bm \gamma}_m,\mathcal{H}_{i-1}\right)\right]
\cdot \frac{\sum_{m=1}^\infty p_{m,i}\mathbb{E}_{{\bm \tau}_{m,i}}\left[\hat{R}_{m,i}\left({\bm \gamma}_m,\mathcal{H}_{i-1}\right)\right]}{\sum_{m=1}^\infty p_{m,i}\mathbb{E}_{{\bm \tau}_{m,i}}\left[\hat{x}_{k,m,i}\left({\bm \gamma}_m,\mathcal{H}_{i-1}\right)\right]} 
\cdot \mathbbm{1}_{i\leq N_T}\Big|\mathcal{H}_{i-1}\right]}{\mathbb{E}\left[\sum_{i=1}^\infty x_{k,i}\mathbbm{1}_{i\leq N_T}\right]} \\
&\geq\frac{\sum_{i=1}^\infty \mathbb{E}_{\mathcal{H}_{i-1}}\left[ \sum_{m=1}^\infty p_{m,i} \mathbb{E}_{\tau_i}\left[\hat{x}_{k,m,i}\left(\gamma,\mathcal{H}_{i-1}\right)\right]
\cdot R^*\left(\mathcal{H}_{i-1}\right)
\cdot \mathbbm{1}_{i\leq N_T}\Big|\mathcal{H}_{i-1}\right]}{\mathbb{E}\left[\sum_{i=1}^\infty x_{k,i}\mathbbm{1}_{i\leq N_T}\right]} \\
&\geq R_{\min},
\end{align}
where (\ref{eq_inc_pf_ren_2}) follows from (\ref{eq_inc_pf_ren_1}) and the monotone convergence theorem, together with the fact that $\mathbb{E}\left[R_i\right]=\sum_{m=1}^\infty p_{m,i}\mathbb{E}\left[R_{m,i}\right]$; $R^*\left(\mathcal{H}_{i-1}\right)$ is the minimum value of $\frac{\sum_{m=1}^\infty p_{m,i}\mathbb{E}_{{\bm \tau}_{m,i}}\left[\hat{R}_{m,i}\left({\bm \gamma}_m,\mathcal{H}_{i-1}\right)\right]}{\sum_{m=1}^\infty p_{m,i}\mathbb{E}_{{\bm \tau}_{m,i}}\left[\hat{x}_{k,m,i}\left({\bm \gamma}_m,\mathcal{H}_{i-1}\right)\right]}$; and $R_{\min}$ is the minimum value of $R^*\left(\mathcal{H}_{i-1}\right)$ over all possible epochs and their corresponding histories, i.e., the minimum over all $i$ and $\mathcal{H}_{i-1}$. This, together with the fact that $\mathbb{E}\left[x_{k,i}\right]=\sum_{m=1}^\infty p_{m,i}\mathbb{E}\left[x_{k,m,i}\right]$, gives the last inequality.

Observe that a policy achieving $R^*\left(\mathcal{H}_{i-1}\right)$ is a policy which is a function of the possible energy arrival patterns in the $i$th epoch ${\bm \tau}_{m,i}$'s only, since the history $\mathcal{H}_{i-1}$ is fixed. Since the energy arrival process is Poisson with rate $1$, it follows that the random vector ${\bm \tau}_{m,i}$ consists of i.i.d. exponential random variables with parameter $1$, and that $\{{\bm \tau}_{m,i}\}$ are also independent across epochs. Therefore, if we repeat the policy that achieves $R_{\min}$ over all epochs, we get a renewal policy where the epoch lengths are also i.i.d., and $\{l_i\}$ forms a renewal process. This completes the proof.

\subsection{Useful Derivatives} \label{apndx_func_der}

In this appendix, we summarize some useful results that we use in the derivation of the optimal threshold policy for problem (\ref{opt_lmda_inc}), which mainly arise while taking derivatives of the Lagrangian constructed for this problem in (\ref{eq_lagrange_inc}).

Focusing on $B=4$, we first start by stating the derivative of the following nested integrals with respect to $y_4(t)$, for some differentiable function $f$:
\begin{align}
I_f\triangleq&\int\limits_{[a,~y_1(\tau_1)-\tau_1,~y_2(\tau_1+\tau_2)-(\tau_1+\tau_2),~y_3(\tau_1+\tau_2+\tau_3)-(\tau_1+\tau_2+\tau_3)]} \hspace{-1in} f\left(y_4(\tau_1+\tau_2+\tau_3+\tau_4)\right) d{\bm \tau}_1^4 \nonumber \\
=&\int\limits_{[a,~y_1(\tau_1)-\tau_1,~y_2(\tau_1+\tau_2)-(\tau_1+\tau_2)]} \int_{u=\tau_1+\tau_2+\tau_3}^{y_3(\tau_1+\tau_2+\tau_3)}f\left(y_4(u)\right)du d{\bm \tau}_1^3, \nonumber \\
\end{align}
where the second equality follows by the change of variables: $u=\tau_1+\tau_2+\tau_3+\tau_4$. This gives
\begin{align}
\frac{\partial I_f}{\partial y_4(t)}= \int\limits_{\substack{[a,~y_1(\tau_1)-\tau_1,~y_2(\tau_1+\tau_2)-(\tau_1+\tau_2)] \\ \tau_1+\tau_2+\tau_3\leq t}} d{\bm \tau}_1^3 f^{\prime}(y_4(t)) \triangleq m_3(a,y_1,y_2,t)f^{\prime}(y_4(t)),
\end{align}
where $f^\prime$ is the derivative of the function $f$.

Next, we mention a couple of results that have to do with taking derivative of the above nested integrals, $I_f$, with respect to boundary limits. The first is when we take derivative with respect to the function in the inner most boundary limit, $y_3(t)$. Towards that end, we apply the change of variables $u=\tau_1+\tau_2+\tau_3$ and rewrite $I_f$ as follows:
\begin{align}
I_f=\int\limits_{[a,~y_1(\tau_1)-\tau_1]} \int_{u=\tau_1+\tau_2}^{y_2(\tau_1+\tau_2)} \int_{\tau_4=0}^{y_3(u)-u} f\left(y_4\left(u+\tau_4\right)\right) d\tau_4 du d{\bm \tau}_1^2.
\end{align}
Now using Leibniz's rule, we get
\begin{align}
\frac{\partial I_f}{\partial y_3(t)}= \int\limits_{\substack{[a,~y_1(\tau_1)-\tau_1] \\ \tau_1+\tau_2\leq t}} d{\bm \tau}_1^2 f\left(y_4\left(y_3(t)\right)^-\right) \triangleq m_2(a,y_1,t)f\left(y_4\left(y_3(t)\right)^-\right).
\end{align}

Finally, we discuss the case when we take derivative with respect to a function in one of the boundary limits in the mid integrals, $y_k(t)$, $k=1,2$. For $k=2$, similar to what we did above, we apply the change of variables $u=\tau_1+\tau_2$, and rewrite $I_f$ as follows:
\begin{align}
I_f=\int_{\tau_1=0}^a\int_{u=\tau_1}^{y_1(\tau_1)}\int\limits_{[y_2(u)-u,~y_3(u+\tau_3)-(u+\tau_3)]} f\left(y_4(u+\tau_3+\tau_4)\right) d\tau_1dud{\bm \tau}_3^4,
\end{align}
which, upon using Leibniz rule, gives
\begin{align}
\frac{\partial I_f}{\partial y_2(t)}= &\int_{\substack{\tau_1=0 \\ \tau_1\leq t}}^a d\tau_1 \int_{\tau_4=0}^{y_3(y_2(t)^-)-y_2(t)} f\left(y_4\left(y_2(t)^-+\tau_4\right) \right) d\tau_4 \nonumber \\
\triangleq &m_1(a,t) \int_{\tau_4=0}^{y_3(y_2(t)^-)-y_2(t)} f\left(y_4\left(y_2(t)^-+\tau_4\right) \right) d\tau_4.
\end{align}
For $k=1$, we do not need a change of variables; we directly get
\begin{align}
\frac{\partial I_f}{\partial y_1(t)}= \int\limits_{[y_2(y_1(t)^-)-y_1(t)^-,~y_3(y_1(t)^-+\tau_3)-(y_1(t)^-+\tau_3)]} \hspace{-1in} f\left(y_4(y_1(t)^-+\tau_3+\tau_4)\right) d{\bm \tau}_3^4.
\end{align}

%============

\end{document}